\documentclass[11pt]{article}
\usepackage{natbib,amsmath,amsthm,amssymb,color,graphicx}
\usepackage{times}
\usepackage{epsfig}

\usepackage[usenames,dvipsnames]{xcolor}
\newcommand{\comment}[1]{{#1}}  

\newcommand{\g}[1]{{#1}}

\newtheorem{alg}{Algorithm}

\newcommand{\thc}{\vartheta} 
\newcommand{\thmod}{{\mathbf{\boldsymbol{\thc}}}} 

\newcommand{\piv}{{\mathbf{\boldsymbol{\pi}}}}

\newcommand{\etav}{{\boldsymbol{\eta}}} 
\newcommand{\xiv}{\boldsymbol{\xi}} 
\newcommand{\Siv}{{\mathbf{S}}}

\newcommand{\thetm}{\theta} 
\newcommand{\thetav}{{\mathbf{\boldsymbol{\thetm}}}} 

\newcommand{\betac}{\beta} 
\newcommand{\betar}{\boldsymbol{\betac}} 
\newcommand{\betav}{\betar} 

\newcommand{\pl}{\pi}  
\newcommand{\plv}{{\mathbf{\boldsymbol{\pl}}}}  
\newcommand{\plswv}[1]{\plv_{#1}} 
\newcommand{\plsw}[2]{\pl_{#1,#2}} 
\newcommand{\catd}{D}

\newcommand{\alphaf}{\alpha} 
\newcommand{\alphav}{\boldsymbol{\alphaf}} 

\newcommand{\im}[1]{^{(#1)}}

\newcommand{\Ew}[1]{\mbox{\rm E}(#1)}   

\newcommand{\Gamfun}[1]{\Gamma (#1)}

\newcommand{\iid }{\mbox{\rm i.i.d.}}

\newcommand{\indic}[1]{I_{\{#1\}}}

\newcommand{\mean}[1]{\overline{#1}}

\newcommand{\Probsym}{\mbox{\rm Pr}}
\newcommand{\Prob}[1]{\Probsym (#1)}

\newcommand{\rvY}{Y} 
\newcommand{\rvYm}{{\mathbf \rvY}} 

\newcommand{\yc}{y}
\newcommand{\ym}{{\mathbf \yc}} 
\newcommand{\ymd}{r} 

\newcommand{\pdf}[3]{f_{   #1 }(#2;#3)} 

\newcommand{\Betadis}[1]{\mathcal{B}\left(#1\right)}

\newcommand{\Dir}[1]{ \mathcal{D}\left(#1\right)}

\newcommand{\DP}[1]{ \mathcal{DP}\left(#1\right)}

\newcommand{\Gammad}[1]{ \mathcal{G}\left(#1\right)}

\newcommand{\Mulnom}[1]{\mbox{\rm MulNom}\left(#1\right)}

 \newcommand{\Normal}[1]{ \mathcal{N}\left(#1\right)}

\newcommand{\Poi}[1]{\mathcal{P}\left(#1\right)}

\newcommand{\Poipdfa}[2]{\pdf{P}{#1}{#2}}

\newcommand{\Uniform}[1]{\mathcal{U}\left[#1\right]}

\newcommand{\eye}{\textsc{Eye Tracking Data}}
\newcommand{\fault}{\textsc{Fabric Fault Data}}
\newcommand{\fear}{\textsc{Childrens' Fear Data}}

\newcommand{\alz}{\textsc{Alzheimer Data}}
\newcommand{\DLBCL}{\textsc{DLBCL  Data}}

\newcommand{\GDP}{\mathcal{G}}

\newcommand{\xis}{\xi}                   
\renewcommand{\xiv}{\boldsymbol{\xis}}                   
\newcommand{\omegas}{\omega}             
\newcommand{\Omegas}{\boldsymbol{\Omega}}
\newcommand{\rvXm}{{\mathbf X}} 

\newcommand{\Kni}{\Kn ^{-i}}

\renewcommand{\indic}[1]{I{\{#1\}}}
\newcommand{\Ktrunc}{K}
\newcommand{\parti}{{\cal P}}

\newcommand{\alphaDP}{\alpha}
\newcommand{\ala}{a_\alpha}
\newcommand{\alb}{b_\alpha}

\newcommand{\baseG}{\mathcal{G}_0}

\newcommand{\Dirinv}[2]{\mathcal{D}_{#1}\left(#2\right)}

\newcommand{\etaw}{\eta}

\newcommand{\ed}[1]{e_{#1}} 
\newcommand{\eda}{a_e} 
\newcommand{\edb}{b_e} 

\newcommand{\Ktrue}{K_{tr}}

\newcommand{\Kn}{K _+}
\newcommand{\Knhat}{\hat{K} _+}

\newcommand{\stick}{v}

\newcommand{\sta}{a}
\newcommand{\stb}{b}

\newcommand{\by}{\mathbf{y}}
\newcommand{\bx}{\mathbf{x}}

\newcommand{\cG}{\mathcal{G}}

\newcommand{\xm}{\boldsymbol{x}}
\newcommand{\alpv}{\betav}

\begin{document}

\title{From here to infinity \g{ -- sparse} finite  versus Dirichlet process mixtures  in model-based clustering} 

\author{Sylvia Fr\"uhwirth-Schnatter and Gertraud Malsiner-Walli\\
 Institute for Statistics and Mathematics,\\Vienna University of Economics and Business (WU)  \\
             Email:{\tt sfruehwi@wu.ac.at} and {\tt gmalsine@wu.ac.at}}

\maketitle

\begin{abstract}
 In model-based clustering mixture models are used to group data points into clusters.
   A useful concept  introduced for Gaussian mixtures by Malsiner Walli et al (2016)
  are sparse finite mixtures, where the prior distribution on the weight distribution of a mixture with $K$ components is chosen in such a way that  a
  priori the number of clusters in the data is random and is allowed  to be smaller than $K$ with high probability. The number of clusters is then inferred a posteriori from the data.

 The present paper makes the following contributions in the context of sparse finite mixture modelling. First, it is illustrated that the concept of sparse finite mixture is very generic and easily
extended to cluster various  types of non-Gaussian data, in particular discrete data and continuous multivariate data arising from non-Gaussian clusters.
Second,  sparse finite mixtures are compared  to Dirichlet process mixtures with respect to their ability to identify the number of  clusters.
For both model classes, a random hyper prior is considered for the parameters determining the weight distribution. By suitable
matching of these priors, it is shown that the choice of this  hyper prior is far more influential on the cluster solution than whether a sparse
finite mixture or a Dirichlet
 process mixture is taken into consideration.
 \end{abstract}




\section{Introduction}

 In the present paper,  interest lies in the use of mixture  models to cluster   data  points 
 into groups of similar objects; see \citet{fru-etal:han} for a review of mixture analysis. Following the pioneering papers
 of  \citet{ban-raf:mod} and \citet{ben-etal:inf},   model-based clustering using finite mixture models  has found numerous applications, see  \citet{gru:mod} for a comprehensive review.

 For finite mixtures, the number $K$ of components is an 	unknown, but fixed quantity and the need to specifiy $K$   in advance is considered  one of the major drawbacks of  applying finite mixture models in a clustering context.  Many methods have been suggested to estimate $K$  from the data such as  BIC \citep{ker:con}, marginal likelihoods \citep{fru:est}, or the integrated classification likelihood \citep{bie-etal:ass}, but typically these methods require to fit several finite mixture models with  increasing $K$. Alternatively,  one-sweep methods such as  reversible jump MCMC  \citep{ric-gre:bay,del-pap:mul} have been suggested, 
but are challenging  to  implement.

As an alternative to finite mixtures,  Dirichlet process mixtures  \citep{fer:bay_den,esc-wes:bay}  were applied in a clustering context   by \cite{qui-igl:bay} and \citet{med-etal:bay}, 
among many others. Using a Dirichlet process prior  \citep{fer:bay_ana,fer:pri} for the parameters generating the data points, Dirichlet process mixtures allow infinite components by construction. Posterior inference  focuses on the partitions  and  clusters   induced by the Dirichlet process prior on the data points.  The number of non-empty clusters is random    by construction  and can be inferred from the data using easily implemented Markov chain Monte Carlo samplers, see e.g.~\citet{mue-mit:bay}.

Recently,  the concept of sparse finite mixtures has been introduced within the framework of Bayesian model-based clustering
\citep{mal-etal:mod,mal-etal:ide}  as a bridge between standard finite mixture  and Dirichlet process mixture models.
Based on theoretical results derived by \citet{rou-men:asy},  the  sparse finite mixture approach  relies on specifying a  sparse symmetric Dirichlet prior  $\Dirinv{K}{\ed{0}}$ on the  weight distribution of an overfitting finite mixture distribution, where the number of components  is larger than the number of clusters in the data.  By choosing small values for the hyperpararmeter $\ed{0}$, the sparse Dirichlet prior  is designed to favour weights close to zero.   \citet{mal-etal:ide} investigate  the partitions induced by such a sparse finite mixture model  and show that the corresponding number of  clusters created in the data  is not fixed a priori.  Rather,  as for Dirichlet process mixtures,  it is random by construction  and can be inferred from the data using common Markov chain Monte Carlo methods.

 The present paper makes  two contributions in the context of sparse finite mixture modelling.
  As a first contribution, it is   illustrated that the concept of sparse finite mixtures, which was originally developed  and investigated in the framework of Gaussian
  mixtures, is very generic and can be easily 	extended to cluster
  a broad range  of non-Gaussian data, in particular discrete data and continuous multivariate data arising from non-Gaussian clusters,
see also  \citet{mal-etal:eff}.   As mentioned above, an advantage of sparse finite mixtures is that model selection with respect to the number of clusters is possible  within one-sweep samplers without the need to design sophisticated proposals within trans-dimensional approaches such as  reversible jump MCMC.  Performing model selection   without computer-intensive methods is  of particular interest for mixtures of non-Gaussian components where the calculation of the marginal likelihood can be  cumbersome 
and  almost impossible for large $K$. 
A wide range of applications, including  sparse Poisson mixtures, sparse  mixtures of generalised linear models  for  count data,
  and   sparse latent class models for multivariate categorical data, demonstrate that sparse finite mixtures provide a useful method  for selecting the number of clusters for such data.
	
 A second aim of the paper is to compare  sparse finite mixtures to Dirichlet process mixtures with respect to their ability to identify the number of  clusters.
As shown by \citet{gre-ric:mod},  a $K$ component finite mixture model with  symmetric Dirichlet prior  $\Dirinv{K}{\alphaDP/K}$
on the weights approximates a Dirichlet process mixture with concentration parameter  $\alphaDP$ 
as $K$ increases.
For  $\alphaDP$ given, this sequence of finite mixtures increasingly  becomes sparse,
as $\ed{0}=\alphaDP/K$ decreases with increasing $K$ and the Dirichlet process mixture can be seen  as the limiting case of a sparse finite mixture
with   $K=\infty$.
	Both for sparse finite mixtures  and Dirichlet process mixtures,   the number  of non-empty clusters is random a priori  and can be estimated from the data. Since Dirichlet process  mixtures  can be  inconsistent with respect to the number of components  \citep{mil-har:sim},  sparse finite mixtures appear to be an attractive  alternative which shares many interesting features with  Dirichlet process mixtures.

Finite mixture  and  Dirichlet process mixture  models  are generally considered to be quite different approaches. \g{Irrespectively of this, } the aim of the paper is not to discuss pros and cons of the two model classes. Rather,  it will be shown that  both  model classes  yield  similar inference with respect to the number of clusters,  once  the hyper prior for 
 $\alphaDP$  is  matched to hyper priors on $\ed{0}$ that induces sparsity.
Comparisons between  sparse finite mixtures and Dirichlet process mixtures in applications based on  Poisson mixtures,  mixtures of generalised linear models,   and   latent class models  illustrate that the choice of the hyper prior on  $\ed{0}$ and $\alphaDP$ is far more influential on the cluster solution than which of the two model classes  is taken into consideration.  


The rest of the paper is organized as follows.
	 	 Section~\ref{sec2}
	 	summarizes the concept of sparse finite mixtures and investigates their relationship to Dirichlet process mixtures. 
	 	Section~\ref{section4} reviews various finite mixture models with non-Gaussian components.
	 	Section ~\ref{sec:sim}  contains an extensive simulation study where the performance of sparse finite mixtures and Dirichlet process mixtures  in regard to model selection and clustering behavior is investigated in detail for latent class models.
	 	 In Section~\ref{sec:appl},  the sparse finite mixture approach  is illustrated and compared to Dirichlet process mixtures
through case studies for each type of  non-Gaussian mixture model discussed in Section~\ref{section4}.  Section~\ref{sec:dis} concludes with a final discussion of the sparsity prior of the weight distribution in sparse finite mixtures.

\section{From here to infinity}\label{sec2}

\subsection{From finite mixture  distributions to sparse finite mixture models}\label{sec:Model}


The starting point of model-based clustering is a  finite mixture distribution  defined as:
\begin{eqnarray}  \label{mix:dist}
 p(\ym| \thetav_1, \ldots, \thetav_K,\etav)= \sum_{k=1}^K  \etaw_k f_{\cal T}(\ym|\thetav_k),
\end{eqnarray}
where  the  component densities
 $f_{\cal T}(\ym|\thetav_k)$  arise  from the same distribution family ${\cal T}(\thetav)$, each  with  weight $\etaw_k$, and
 $\sum_{k=1}^K  \etaw_k=1$.
 Data $\ym $  generated  from such a mixture distribution can be univariate or multivariate, continuous, discrete-valued or  mixed-type,
outcomes of a regression model, or even time series data;   see \citet{fru:book} for a comprehensive review of finite mixture distributions.  

%
 Clustering arises in a natural way for an  \iid\ sample  
  from the finite
 mixture distribution (\ref{mix:dist}), since each observation $\ym_i$  can be associated with the component, indexed by $S_i$, that generated this data point:
  \begin{eqnarray} \label{hiergibb1}
&S_i | \etav  \sim  \Mulnom{1;\eta_1,\ldots,\eta_K},  & \\
&  \ym_i|S_i \sim {\cal T}(\thetav_{S_i}). & \nonumber
\end{eqnarray}
If  $N$ \iid\ data points $\ym_1,\ldots,\ym_N$  are drawn from the finite mixture distribution (\ref{mix:dist}),
then the sequence $\Siv=(S_1,\ldots,S_N)$ is the collection of all component  indicators that were used
to generate the data.
Obviously,  $\Siv$ defines a partition $\parti$  of the data.
Let    $N_k $
  be the number of observations generated by  component  $k$, $k=1,\ldots,K$.
Then  (\ref{hiergibb1}) implies that:
 \begin{eqnarray} \label{knmul}
N_1  ,\ldots, N_{K}  | \etav  \sim  \Mulnom{N;\eta_1,\ldots,\eta_K}.
\end{eqnarray}
Depending on the weight distribution  $\etav=(\eta_1,\ldots,\eta_K)$ appearing in (\ref{mix:dist}),
  multinomial sampling according to (\ref{knmul})
 may lead to partitions with $N_k=0$. 
  In this case, 
  fewer than $K$  mixture components were used to generate the $N$ data points which
contain $K_+$ data clusters, i.e.
\begin{eqnarray}  \label{KnSiv}
\Kn= K - \sum_{k=1}^K  \indic{N_k=0}.
\end{eqnarray}
%
%
It is important to realize that in model-based clustering
interest lies foremost in estimating  the number of clusters 
in the  data, rather than the number of    components  
 of the mixture distribution (\ref{mix:dist}).
Hence, in model-based clustering based  on finite mixtures,  it is extremely important  to distinguish between  the
order $K$ of  the underlying mixture distribution  and   the number of  (non-empty) clusters   $\Kn$ in the $N$  data points.
For finite mixtures this  difference between $K$ and   $\Kn$  is rarely addressed explicitly,   exceptions
being  \citet{nob:pos}   and, more recently, \citet{mil-har:mix} and \citet{mal-etal:ide}.

\begin{figure}[t]
	\begin{center}
	\includegraphics[width=0.25\textwidth]{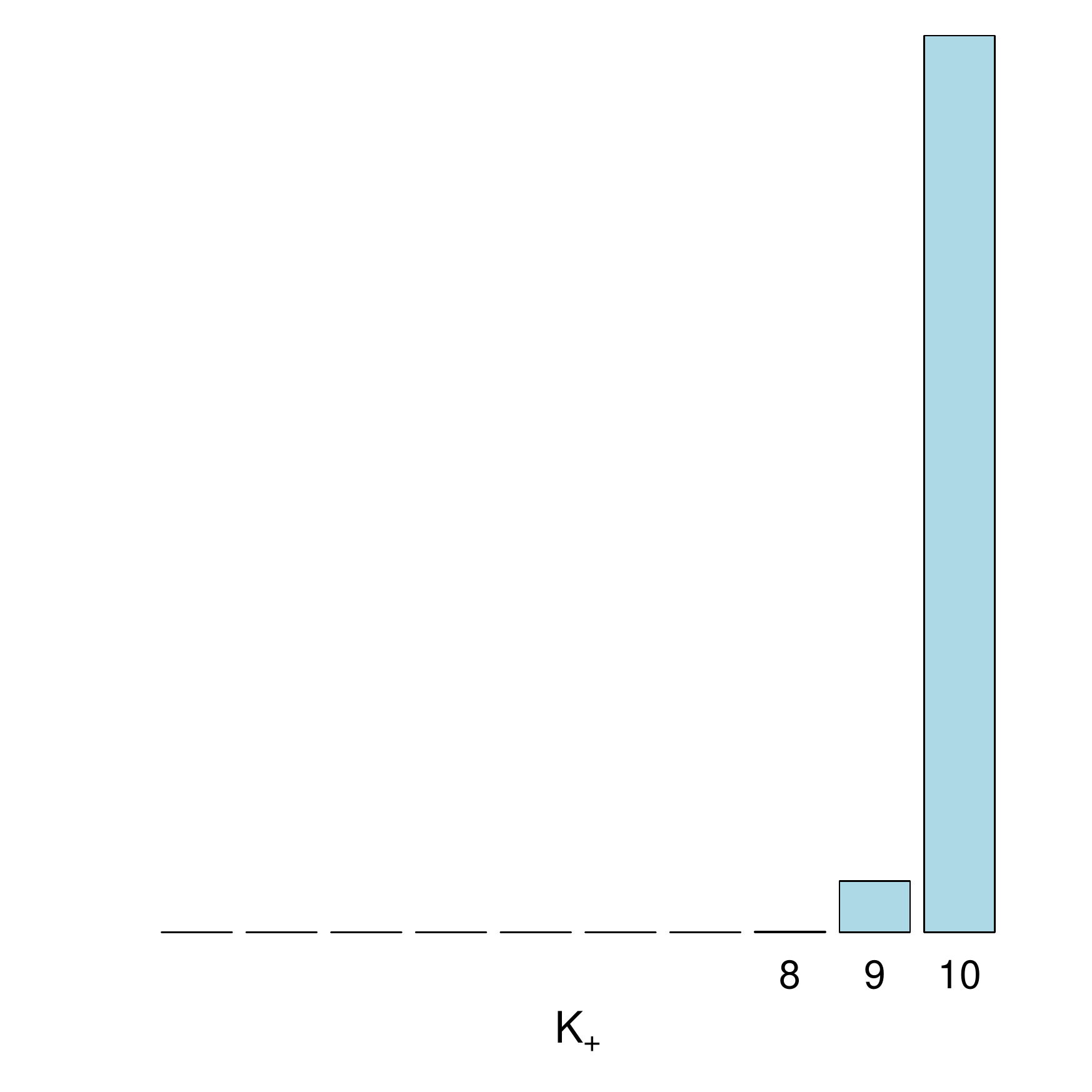}
	\includegraphics[width=0.25\textwidth]{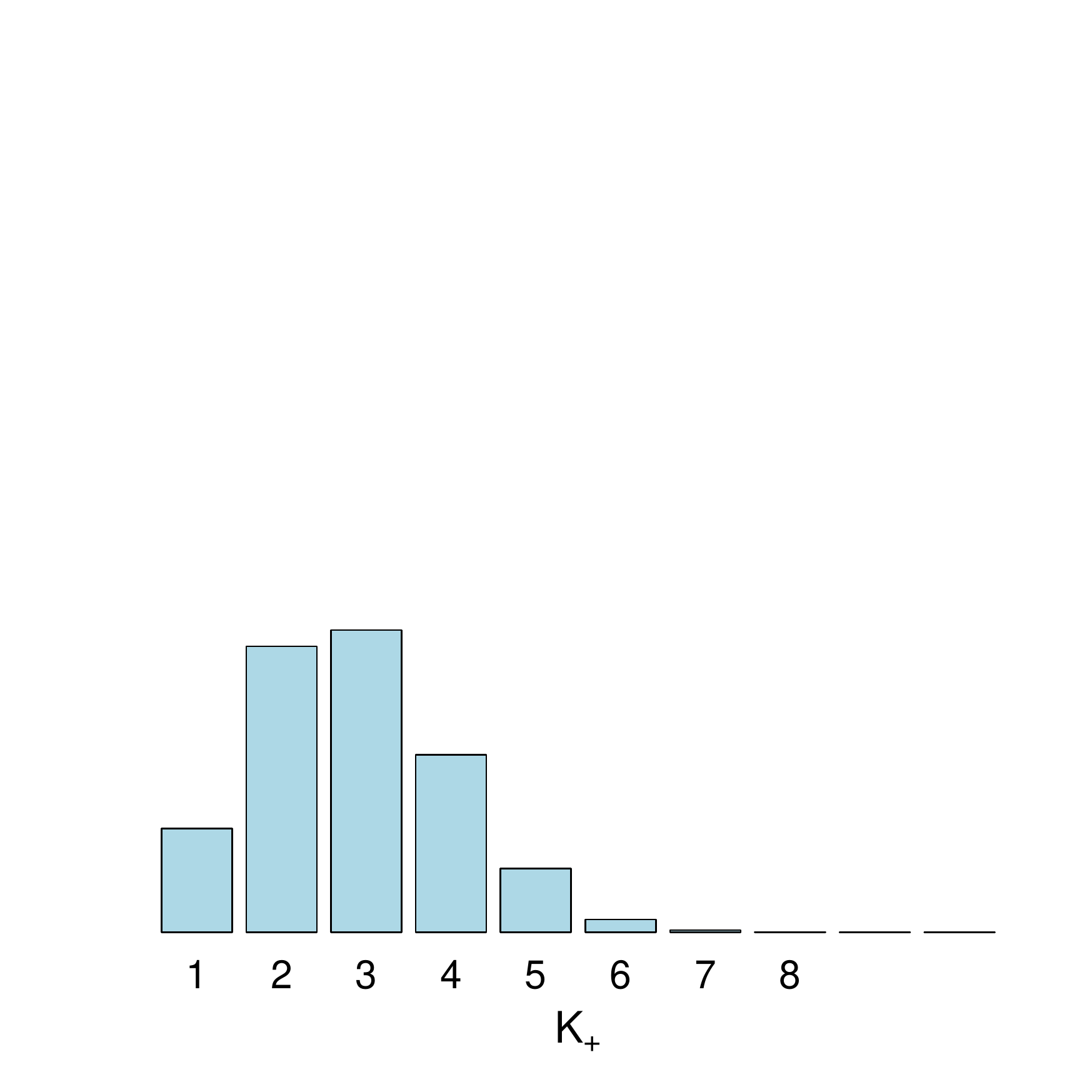}
	\includegraphics[width=0.25\textwidth]{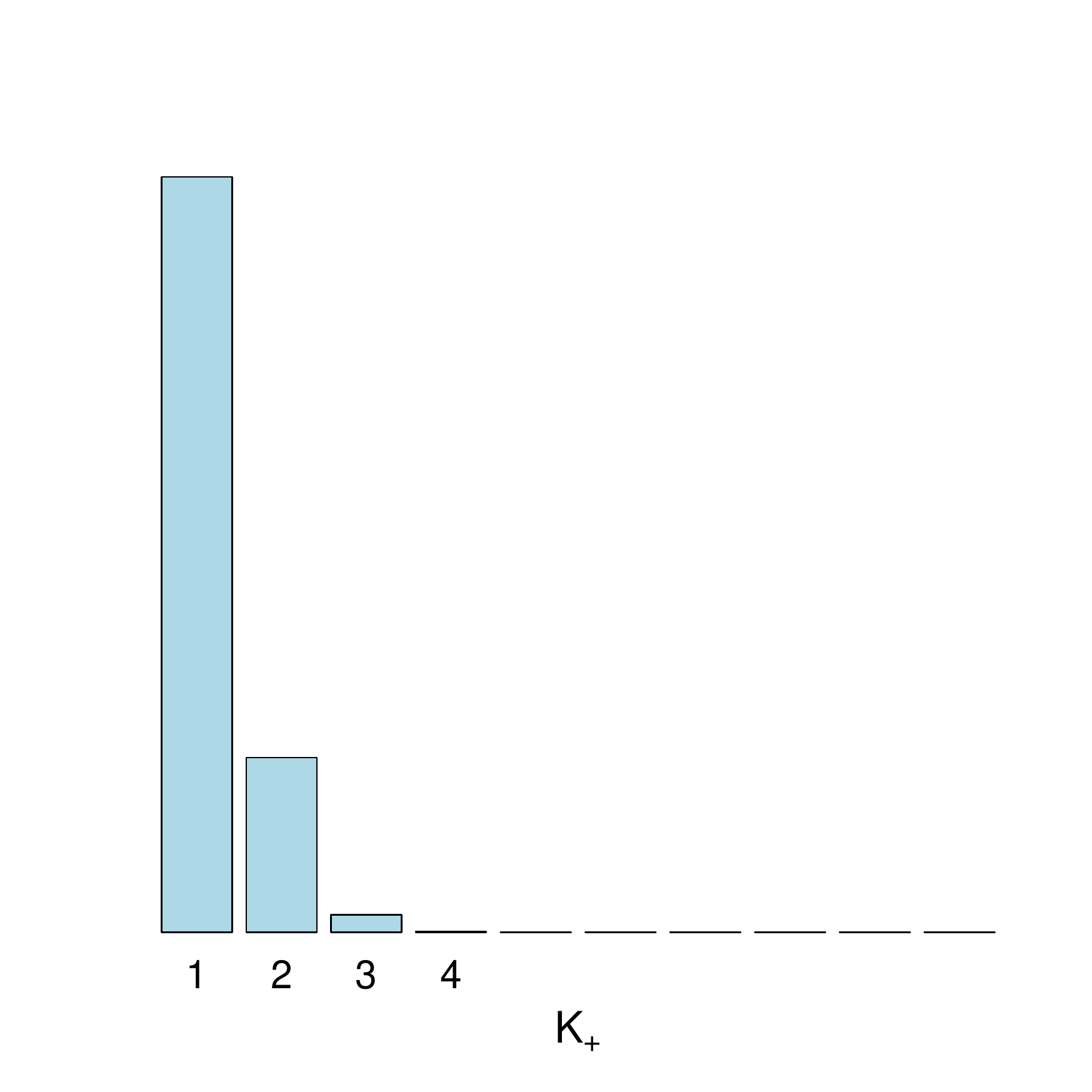}\\
	\includegraphics[width=0.25\textwidth]{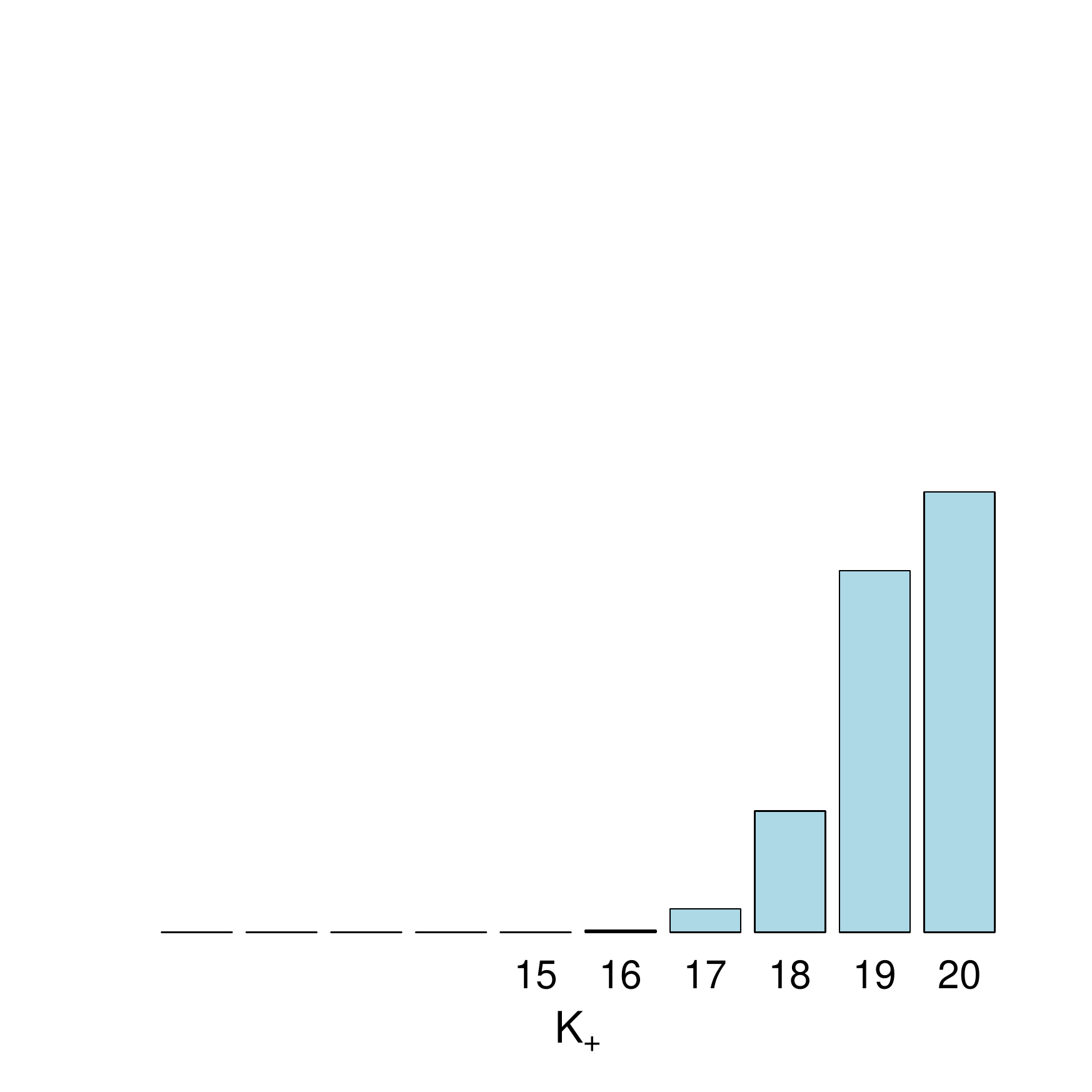}
	\includegraphics[width=0.25\textwidth]{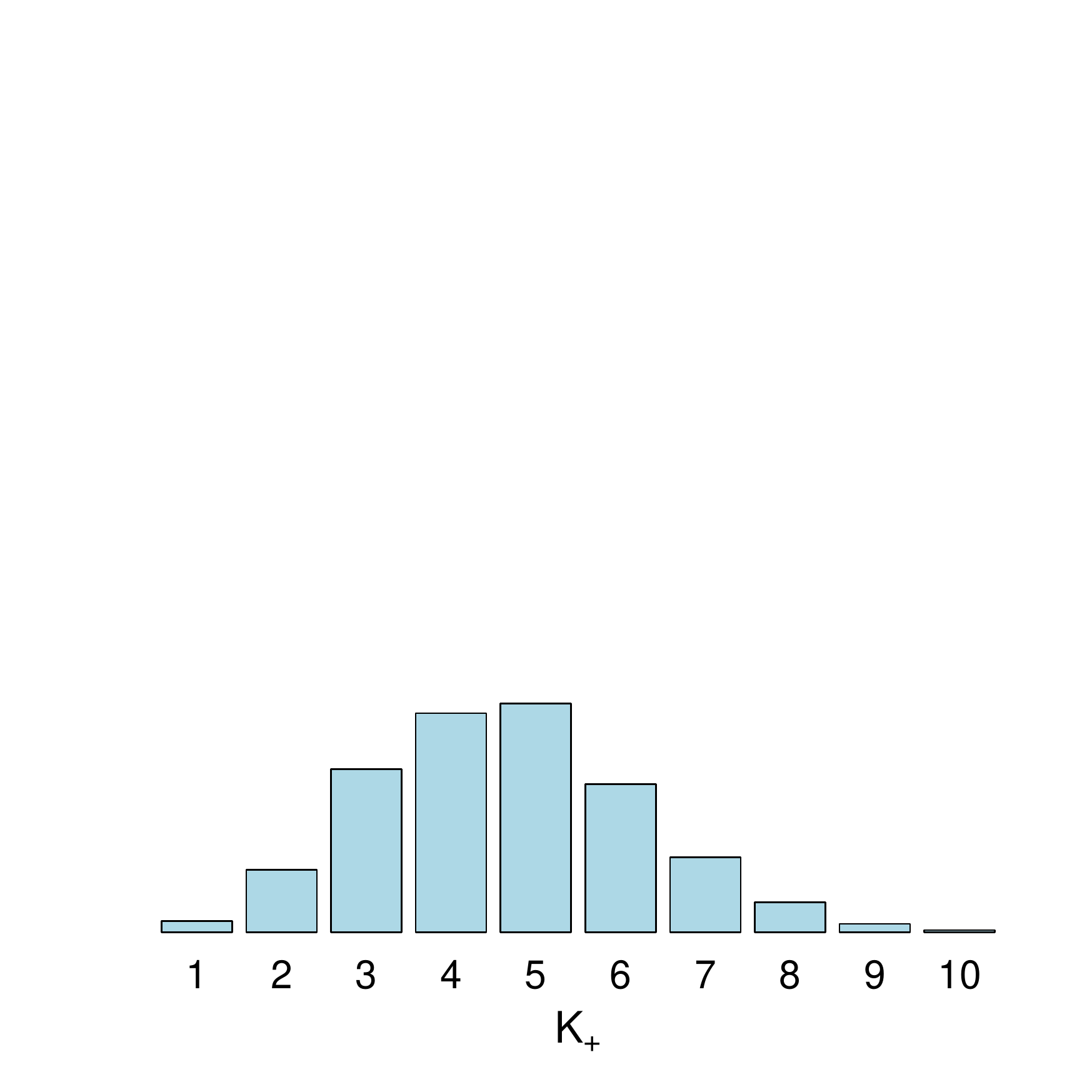}
	\includegraphics[width=0.25\textwidth]{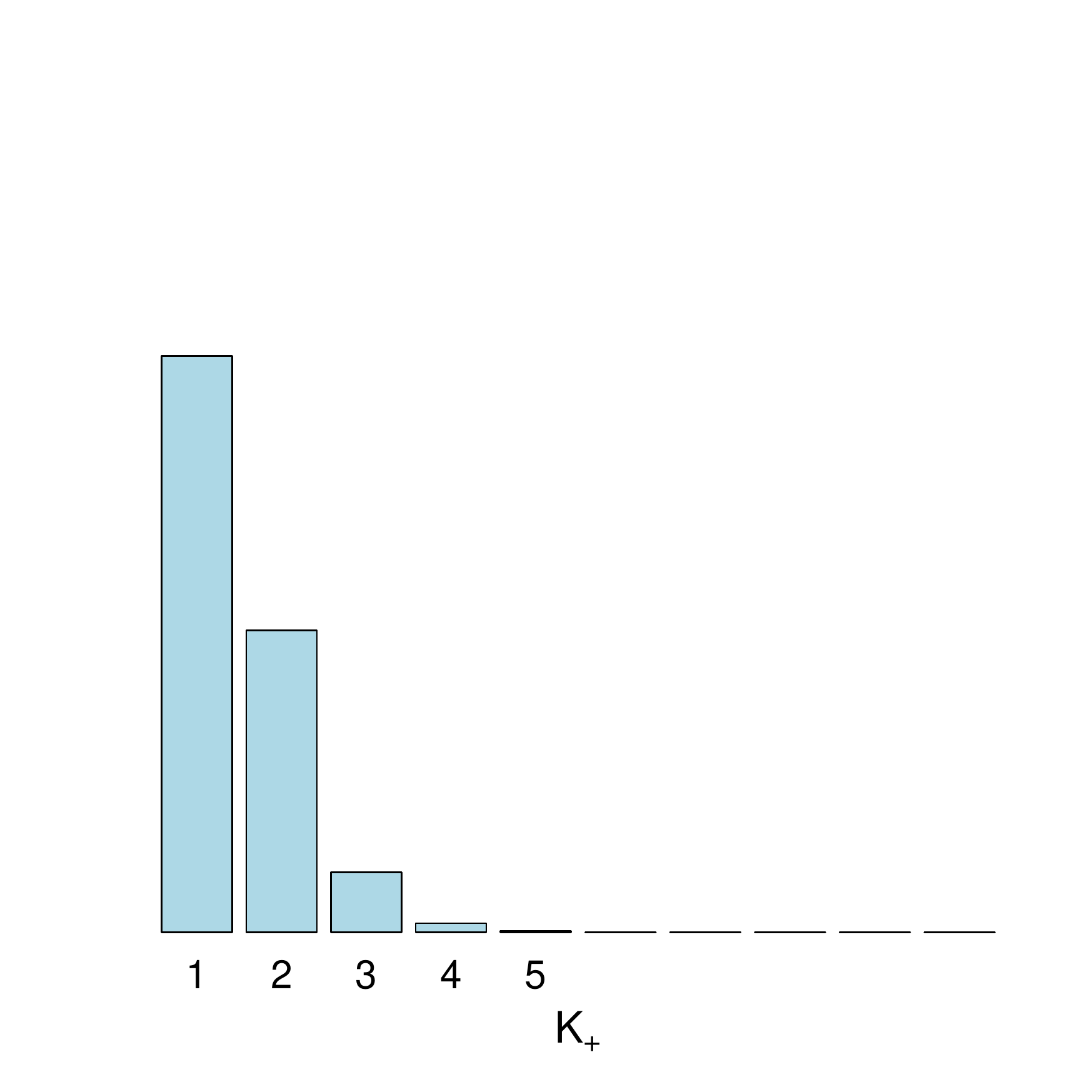}
	\caption{Prior distribution $p(K_+|e_0,K)$  of the  number of data clusters $K_+$ for $N=100$
with  $K=10$ (top row) and $K=20$ (bottom row) and $e_0=4$ (left-hand side), $e_0=0.05$ (middle), and $e_0=0.005$ (right-hand side).}
	\label{plot:priorKplus}
	\end{center}
\end{figure}

   If  finite mixtures are used to  cluster  data  with  the number  of clusters  $\Kn$ being unknown, then it makes sense to choose a prior on the  weight distribution  $\etav=(\eta_1,\ldots,\eta_K)$  that allows a priori that  $\Kn  < K$  with high probability.
This is the very idea of the sparse finite mixture approach  introduced by \citet{mal-etal:mod} for mixtures
of univariate and multivariate   Gaussian  distributions.
 Sparse finite mixture models make a clear distinction between  $K$, the order of the mixture distribution,  and $\Kn$, the number of clusters in the data.

The sparse finite mixture approach  pursues the following idea: if we choose a
mixture model that is overfitting,  then $\Kn < K$  clusters will be present in the data.
Then, as an intrinsically Bayesian approach, for a given value of $K$ a
prior distribution on $\Kn$ is  imposed which allows  $\Kn$ to be a random variable a priori, taking values smaller than  $K$ with high probability.
 This is achieved in an indirect way through choosing an appropriate prior on the weight distribution   $\etav=(\eta_1,\ldots,\eta_K)$,
the commonly used   prior being the  Dirichlet  distribution  $\etav \sim \Dir{\ed{1}, \ldots, \ed{K}}$. Very  often,
 a symmetric Dirichlet prior is assumed with $\ed{k} \equiv \ed{0}$, $k=1,\ldots,K$; such a prior will be denoted by
$\etav \sim \Dirinv{K}{\ed{0}}$.
If  $e_0$ is a  small value, then many of the $K$  weights  will be small a priori, implying that not all $K$
components will  generate a cluster of their own and, according to  (\ref{knmul}),  $\Kn<K$ with high probability.
The prior of  $K_+$ depends on both $e_0$  and $K$, as illustrated  in Figure \ref{plot:priorKplus}, showing the prior distribution
$p(K_+|e_0,K)$  for various values of  $K$ and $e_0$. For increasing $K$ and $e_0$ also the expected number of clusters $K_+$ increases.


Given data  $\ym=(\ym_1, \ldots, \ym_N)$, the  posterior distribution  $p(K_+|\by)$   
of  $K_+$  is used to estimate the number of data clusters.
For each iteration $m$  of   MCMC sampling (to be discussed in  Subsection~\ref{bay_est}),  a partition $\Siv \im{m}$ is sampled and given  the corresponding
occupation numbers $N_1^{(m)}, \ldots,  N_K^{(m)}$,  the number of non-empty clusters $K_+^{(m)}$ is  determined  using (\ref{KnSiv}).
Then,  $\hat{K}_+$ is  estimated  by the most frequent number of non-empty components: $\hat{K}_+=\text{mode}\{p(K_+|\by)\}$.

\begin{figure}[h]
	\begin{center}
		\includegraphics[width=0.27\textwidth]{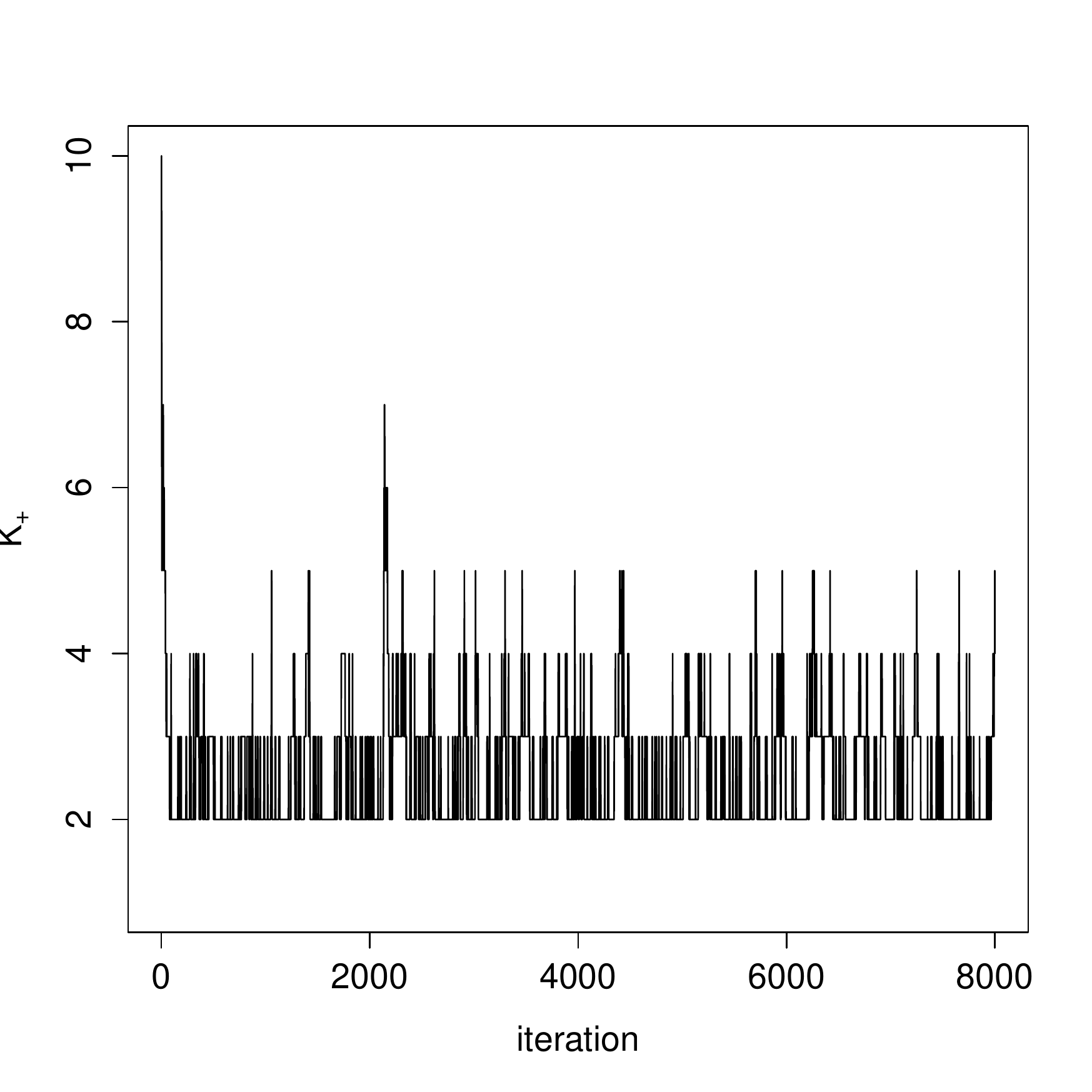}
		\includegraphics[width=0.27\textwidth]{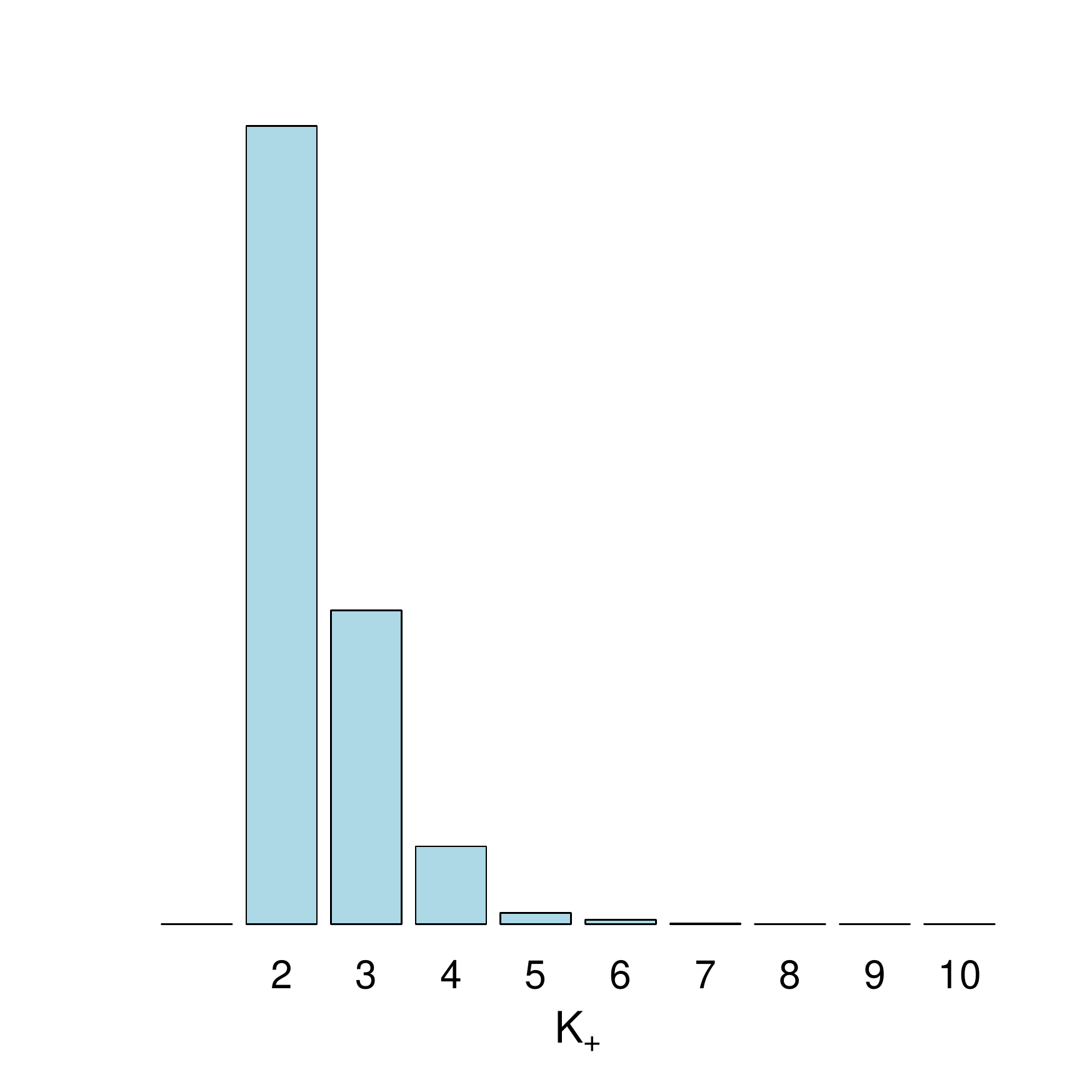}
	
		\caption{\fear ;~trace plot of the number of   clusters $K_+$ during MCMC sampling (left-hand side) and posterior distribution $p(K_+|\by)$ after removing the burn-in  (right-hand side).}
		\label{plot:feartrace}
	\end{center}
\end{figure}

To illustrate the practical procedure, a sparse latent class model with $K=10$ and $e_0=0.005$ is fitted to the \fear\   which will be investigated
in Subsection~\ref{secfear}. In Figure \ref{plot:feartrace}, the corresponding trace plot of  $K_+^{(m)}$  is plotted for  8000 MCMC  iterations. Whereas
 observations are assigned to all 10 components at the very beginning, most components become empty rather quickly and the chain switches between 2 and 5
  nonempty components in its steady state. With the  mode of the posterior $p(K_+|\by)$ being clearly equal to two,
two data clusters are estimated for this data set .

\subsection{From sparse finite mixture models to Dirichlet process mixtures}  \label{refernces}

Sparse finite mixture models  allow to estimate the number $\Kn$  clusters   a posteriori, given the data.
A sparse finite mixture  is \lq\lq sparse\rq\rq\  insofar, as it uses less than $K$ components of the underlying finite mixture distribution
for clustering the data.
In this sense, the sparse finite mixture approach  is related to Bayesian non-parametric  approaches
 such as Dirichlet process mixtures   (DPM)  based on the Dirichlet process prior $\GDP \sim \DP{\alphaDP,\baseG}$
  with concentration parameter $\alphaDP$ and base measure $\baseG$.

 Random probability measure priors like the Dirichlet process prior  lead to countably   infinite mixtures,
 which have a representation similar to (\ref{mix:dist}), however
  with $K=\infty$:
 \begin{eqnarray*}  \label{eqsec11}
p(\ym)= \int  f_{\cal T}(\ym|\thetav) \GDP (d \thetav)=
   \sum_{k=1}^{ \infty}  \etaw_k    f_{\cal T}(\ym|\thetav_k),
\end{eqnarray*}
where $\etaw_k$ are random weights such that $\sum_{k=1}^\infty  \etaw_k=1$ almost surely.
 With $K$ being infinite,  the focus  of DPM automatically lies  on the partitions implied by the Dirichlet process prior  and
the corresponding number  of clusters  $\Kn$. In this sense, DPM implicitly make a distinction between $K$ and $K_+$.

 If the base measure $\thetav_k \sim \baseG$ of a DPM is the same as the prior  $p(\thetav_k )$ in a finite mixture model, then
 the only difference between  these two model classes  lies in the
prior on the weight distribution.  A stick-breaking representation \citep{set:con} of the weights  $\etaw_1, \etaw_2, \etaw_3, \ldots $
  in terms of a sequence $\stick_1, \stick_2, \stick_3, \ldots $ of  independent random   variables,  so-called sticks,
    allows to construct the weights iteratively  for both model classes:
   \begin{eqnarray} \label{stick:weisti}
    \etaw_1=\stick_1, \quad   \etaw_2=(1-\stick_1)\stick_2,  \qquad  \etaw_k= \stick_k \prod_{j=1}^{k-1} (1-  \stick_j ), \quad \nu_k\sim  \Betadis{a_k,b_k}. 
\end{eqnarray}
 However, the two model classes   differ in the parameters $a_k$ and $b_k$, as    $\stick_k \sim \Betadis{1,\alphaDP}, k=1,2,\ldots$,
 for  a DPM with precision parameter $\alphaDP$ and   $\stick_k \sim \Betadis{\ed{0},(K-k)\ed{0}}, k=1,\ldots, K-1$,  $\stick_K=1$  for a
  finite mixture model with parameter $e_0$,   see e.g.~\citet{fru:dea}.

To understand the clustering behavior of both model classes, it is illuminating to compare  them  in regard to
the prior probability to create a new cluster when reallocating an observation $\by_i$, given all remaining observations $\by_{-i}$.
For a DPM  this probability  is equal to \citep{lau-gre:bay}:
 \begin{eqnarray}
{ \frac{   \alphaDP}{  N-1 +   \alphaDP} },  \label{pnewinf}
\end{eqnarray}
 independently of  the current number of non-empty clusters $\Kni$ implied by $\Siv_{-i}$,  where  $\Siv_{-i}$ denotes all indicators  excluding $S_i$.
This leads to  well-known issues with model-based clustering based on DPM. Since  the number of cluster $\Kn \sim \alphaDP \log (N)$
   increases with $N$, \g{it is very likely that   one big cluster
   is found, the sizes of further clusters are geometrically decaying, and many singleton clusters are
   estimated \citep{mue-mit:bay}.}

In contrast, for sparse finite mixtures the probability that observation $\ym_i$ is assigned to an empty cluster, given the
indicators   $\Siv_{-i}$ for all remaining observations,   reads   \citep{lau-gre:bay}:
\begin{eqnarray}
{ \frac{ { \ed{0}  (K-\Kni)}}{  N-1 + \ed{0} K} },  \label{pnew}
\end{eqnarray}
i.e.~the probability to create a new cluster goes to zero as the number of non-empty clusters $\Kni$ increases.
Based on (\ref{pnew}),  \citet{mal-etal:ide} argue that  a sparse finite mixture  with  fixed $K$
 provides a two-parameter alternative to  DPM where  $\Kn \leq K $  is finite, even if $N$ goes to infinity.
Hence, DPM are mainly useful  if the modelling assumption is that the number of data clusters increases with increasing data information
as  is the case  e.g in the text mining framework, where the number of topics may increase, if more documents are considered. As opposed to that,
 sparse finite mixtures are mainly useful for applications  where the underlying assumption is that the data arise from a moderate number of clusters, even  if the number of data points $N$ increases. However, it should be remarked that these recommendations are based on theoretical considerations. As we will see in the simulation study  and the applications,   the  clustering performance of both model classes becomes comparable,
  if the priors of the precision parameters  $\alphaDP$ and $\ed{0}$  driving the stick-breaking representation are appropriately matched, as explained in the following subsection.

\subsection{The importance of hyper priors on the precision parameters}  \label{bay_hyp}

It is obvious from the probabilities to create a new cluster given in   (\ref{pnewinf}) and (\ref{pnew})  that the precision parameters
  $\ed{0}$ and  $\alphaDP$  exercise considerable impact on the resulting  clustering.
 For DPM  
 it is common to assume that $\alphaDP$ is unknown,
typically following a Gamma prior:
    \begin{eqnarray*}
  \alphaDP \sim \Gammad{\ala,\alb},    \label{prioraldpp}
\end{eqnarray*}
where $E(\alpha)=\ala/\alb$.
Choosing a large value $\alb$ is particularly relevant, because it encourages  clustering \citep{mue-mit:bay}.
Commonly, the following prior suggested  by \citet{esc-wes:bay} is applied: $\alphaDP \sim \Gammad{2,4}$ with expectation  $E(\alpha)=0.5$.

For finite mixture models,  it is less common to assume that   $\ed{0}$ is an unknown precision parameter to be estimated from the data - rather $\ed{0}$ is typically fixed.
Choosing $\ed{0}=1$, for instance,
leads to a uniform prior over the unit simplex spanned by all possible weight distributions $\eta_1, \ldots, \eta_K$. \citet{fru:book} recommends choosing  $\ed{0}=4$. This
implies that  the  number of clusters $\Kn$ is equal to the   number of components $K$ with high probability,  see again Figure~\ref{plot:priorKplus} which is sensible only if we assume that the data actually contain  $K$  groups.

 For sparse finite mixtures, where  $\Kn $  is unknown a priori and typically smaller than $K$,   the precision parameter $\ed{0}$
 heavily  influences the probability to create a new cluster given in (\ref{pnew}), see also
  Figure \ref{plot:priorKplus}. Hence,
  \citet{mal-etal:mod} suggested to estimate  $\ed{0}$ from the data  using  the following  Gamma prior:
    \begin{eqnarray*}
    \ed{0} \sim \Gammad{\eda ,\edb},    \label{priore0}
\end{eqnarray*}
where \g{$E(e_0)=\eda/\edb$ }is a small number.
\citet{mal-etal:mod}  compared  the clustering results obtained by  putting  a hyper prior on $e_0$ with an analysis where $\ed{0}$ is a fixed,  small value such as $\ed{0}=0.01$
 for sparse finite mixtures of  Gaussian distributions. The results indicated that   it is important to choose values of  $\eda$ and $\edb$ that imply
 strong prior shrinkage of $\ed{0}$ toward 0, see also \citet{van-etal:ove}. As shown in the present paper, such a
  choice of  $\eda$ and $\edb$  is   also  crucial   for more general sparse finite mixture models in the context
 of clustering discrete data and  data with non-Gaussian clusters.
  A further discussion of this issue will be provided  in  Section~\ref{sec:dis}.

 As will be demonstrated in the applications in Section~\ref{sec:appl},
 sparse finite mixtures lead to sensible estimates of the number of clusters and often coincide with the number of components
 selected by marginal likelihoods based on $\ed{0}=4$. As opposed to that DPM tend to overfit the number of clusters,
 as recently shown by \cite{mil-har:sim}. There is an asymptotic explanation for this behaviour,  however, as will be shown,
 for  moderately sized data sets, this behaviour has to be mainly addressed to the influence  of the hyper prior  on  $\alphaDP $.

 Indeed, the asymptotic relationship $\ed{0}  \approx   \alphaDP/K $ between sparse finite mixtures with $K$ components
  and DPM can be exploited  to  match the priors to each others:
    \begin{eqnarray*}
  \alphaDP \sim \Gammad{\eda,\edb/K}  .    \label{priormatch}
\end{eqnarray*}
A  simulation study and various applications will demonstrate that this matching leads to  a \lq\lq sparse\rq\rq\   DPM
that avoids overfitting the number of clusters. On the other hand, if  a  sparse  finite mixture is matched through $\ed{0} \sim \Gammad{\ala ,K \alb}$ to
a DPM  with common priors such as $\ala=2 ,  \alb=4$, then it tends \g{to lose} its ability to find sensible cluster solutions and overestimates the number of clusters as well.


\subsection{Bayesian inference}  \label{bay_est}

Bayesian inference both for sparse finite mixture model as well as the DPM model  is summarized in Algorithm~\ref{alg1}. It is assumed that the base measure $\baseG$ is equal to the prior  distribution $p(\thetav_k)$.    For both model classes, basically the same Gibbs sampling scheme  can be used with model-specific steps for sampling  the  precision parameters $ \ed{0}$ and $\alpha$.
 Bayesian estimation of a sparse finite mixture is a straightforward extension of MCMC estimation of a standard finite mixture \citep[Chapter~3]{fru:book}
  and   requires only one additional step to update  $ \ed{0}$  \citep{mal-etal:mod}.
  Bayesian inference for the DPM model  relies on full conditional MCMC sampling as introduced in \citet{ish-jam:gib}.


    \begin{alg} \label{alg1}
 Choose an initial classification  $\Siv$ and repeat the following steps:
 \begin{enumerate}
  \item[(a)] Sample from  $\thetav_k|\Siv,\ym$ for all $k=1,\ldots, \Ktrunc$:
  \begin{enumerate}
              \item[(a-1)] for all \g{non-empty} components (i.e.~$N_k \neq 0$), sample $\thetav_k$ from the complete-data posterior $p(\thetav_k|\Siv,\ym)$;
              \item[(a-2)] for all empty components (i.e. $N_k=0$), sample $\thetav_k$ from the prior $p(\thetav_k)$.
  \end{enumerate}
  \item[(b)]  Define $\stick_\Ktrunc=1$  and sample the sticks 
 $ \stick_1,\ldots, \stick_{\Ktrunc-1} $  
 independently from the \g{following  Beta }distributions,
  \begin{eqnarray*} \label{stikpost2}
 \stick_k|\Siv \sim \Betadis{\sta_k + N_k  ,\stb_k + \sum_{l=k+1}^{\Ktrunc}  N_l  }, \qquad  k=1, \ldots, \Ktrunc-1. 
\end{eqnarray*}
 Determine  the weights  from the sticks using the stick-breaking representation (\ref{stick:weisti}).
\item[(c)]   Sample $\Siv| \etav, \ym$ by sampling each $S_i$ independently for $i=1, \ldots, N$:
\begin{enumerate}
  \item[(c-1)]  Sample $u_i|S_i \sim \Uniform{0,\xi_{S_i}}$;
  \item[(c-2)] Sample $S_i$ from following discrete distribution:
 \begin{eqnarray*}  \label{dDPsamp}
     \Prob{S_i=k|u_i,\thetav_1,\ldots, \thetav_\Ktrunc,\etav,\ym } \propto  \frac{\indic{u_i< \xi_k}}{\xi_k}\times  \etaw_k  f_{\cal T}(\ym_i|\thetav_k),
\quad   k=1, \ldots, \Ktrunc .
\end{eqnarray*}
  \end{enumerate}
  \item[(d)] Sample the precision  parameters using an MH step:
  \begin{enumerate}
 \item[(d-1)]  For SFM, sample $\ed{0} $  from $p(\ed{0}|\parti,K) \propto  p(\parti|\ed{0},K) p(\ed{0}) $ where
 \begin{eqnarray*}
\displaystyle p(\parti| \ed{0},K) = \frac{K!}{(K-\Kn) !}
 \frac{\Gamfun{K \ed{0}} }{\Gamfun{N+ K \ed{0}}} \prod_{k: N_k  >0 } \frac{\Gamfun{N_k+\ed{0}}}{\Gamfun{\ed{0}}}. \label{ijkkh:ml}
\end{eqnarray*}
    \item[(d-2)] For DPM, sample $\alphaDP$ from $p(\alphaDP|\parti) \propto  p(\parti|\alphaDP) p(\alphaDP) $ where
     \begin{eqnarray*}  \label{priparti}
\displaystyle p(\parti|\alphaDP)=  \alphaDP^{\Kn}  \frac{\Gamfun{\alphaDP} }{ \Gamfun{N+ \alphaDP}} \prod_{k:N_k>0}
 \Gamfun{N_k}.
\end{eqnarray*}
  \end{enumerate}
   \end{enumerate}
\end{alg}


  \noindent By exploiting the stick breaking representation (\ref{stick:weisti}),  sampling the weight distribution in Step~(b) is  unified for both model classes.
     For DPM models,   classification  in Step~(c) is performed  using  slice sampling  \citep{kal-etal:sli}
   with $\xi_k= (1-\kappa)\kappa^{k-1}$, where   $\kappa= 0.8$, to achieve random truncation.    The truncation level $\Ktrunc$  is chosen   such that
$ 1-  \sum_{k=1}^\Ktrunc  \eta_k  <  \min(u_1,\ldots,u_N) $  \citep{pap-rob:ret}.
   For sparse finite mixtures,  $\xi_k \equiv 1$, and no truncation is performed, i.e.  Step~(c-1) is skipped  and   Step~(c-2) is equal to the standard classification step,  since   $\indic{u_i< \xi_k}/{\xi_k}=1$.

 To sample $ \ed{0}$ in Step~(d-1), we use \g{an  MH-algorithm} with a high level of marginalization, where  $ \ed{0}$ is sampled
     from the conditional posterior   $p( \ed{0}|\parti, K)$ given the partition $\parti$ rather than  from  $p( \ed{0}|\etav)$ as in \citet{mal-etal:mod}.
   Special care has to be exercised for shrinkage priors on  $ \ed{0} $ and  $\alphaDP$,
    when implementing the MH-algorithm in Step~(d), since the acceptance rate often involves the evaluation of the Gamma function for very small values, which can
    lead to numerical problems. However, these  problems can be easily avoided by writing 
$ \Gamfun{x}= \Gamfun{1+x}/ x$ for arguments $x$ close to 0.

The fitted models are identified in order to obtain  a final  partition of the data and to  characterize the data clusters.  We employ the post-processing
procedure suggested by \citet{fru:book} (see also \citet{Mix:Fruehwirth2011})  for finite mixtures and successfully applied in many papers, e.g.~\citet{mal-etal:mod}
and \citet{mal-etal:ide}.
 Roughly speaking, the procedure works as follows. First, the number of data clusters $\hat{K}_+$ is estimated by the mode of the posterior $p(K_+|\by)$.
Then for all posterior draws were  $\Kn \im{m} = \hat{K}_+$, the  component-specific parameters  $\thetav_k$,  or some (lower-dimensional) functional  $\varphi(\thetav_k)$,
are  clustered in the point process representation into $\hat{K}_+$ clusters using $k$-means clustering. A
unique labeling of the draws is obtained and used to reorder all draws, including the sampled allocations. The final partition is then determined by
 the maximum a posteriori (MAP) estimate of the relabelled cluster allocations.

  This procedure  is applied to the MCMC output of both  finite and infinite mixture models. An advantage of this procedure is that the final partition and  the cluster-specific parameters can be estimated at the same time.
%

\section{Sparse finite mixture models for non-Gaussian data} \label{section4}

Sparse finite mixture models were introduced  in \citet{mal-etal:mod} in the framework of Gaussian mixture distributions, however, the underlying concept  is very generic and  can be easily applied to  more or less any mixture distribution.
In this section, we consider  various types of  sparse finite mixture models for non-Gaussian data, including
 sparse latent class models for multivariate categorical data (Subsection~\ref{seclcm}),
 sparse Poisson mixtures  for  univariate discrete data (Subsection~\ref{secpoi})  and
 sparse  mixtures of generalised linear models (GLMs) for regression models with count data outcomes (Subsection~\ref{mixGLM}).
Finally, Subsection~\ref{mixskew} considers clustering continuous data with non-Gaussian clusters  using  mixtures of univariate and multivariate
skew normal  and skew-$t$ distributions.  For each of these classes of mixture models, case studies are  provided  in Section~\ref{sec:appl}  where sparse finite mixtures are compared to Dirichlet process mixtures of the same type.

\subsection{Sparse latent class models} \label{seclcm}

First, we consider model-based clustering of multivariate binary or categorical data  $\{\ym_1,
\ldots,\ym_N\}$, where $\ym_i= (y_{i1},\ldots,y_{i \ymd})$ is the realization of an $\ymd$-dimensional
discrete random variable $\rvYm =(\rvY_1, \ldots,\rvY_{\ymd})$. Mixture models for multivariate discrete
data, usually called latent class models, or  latent structure analysis, have long been recognized as a
useful tool in the behavioral and biomedical sciences, as exemplified by  \citet{laz-hen:lat},
\citet{goo:exp} and  \citet{clo-goo:lat}, among many others; see
also \citet[Section~9.5]{fru:book} for a  review.
 In Subsection~\ref{secfear} we will analyse the \fear\   \citep{ste-etal:sta}  using a sparse latent class model.

In latent structure analysis it is assumed that  the entire dependence between the elements $\rvY_1, \ldots,\rvY_{\ymd}$ of $\rvYm$,
which  are the so-called manifest variables,   is caused by a discrete latent variable $S_i$, the so-called latent class. Therefore, conditional on the
latent variable  $S_i$,  the  variables  $\rvY_1, \ldots,\rvY_{\ymd}$,
 are  stochastically independent.   Latent structure analysis is closely related to multivariate mixture modeling, as
marginally     $\rvYm$  follows a multivariate discrete mixture distribution:
\begin{eqnarray*}
p(\ym_i|\thmod) = \sum_{k=1}^K \eta_k \prod_{j=1}^{\ymd} p (y_{ij}
|  \plswv{k,j} ), 
\end{eqnarray*}
where $\plswv{k,j}$ is a parameter modeling the discrete probability distribution of  $\rvY_{j}$
 in   class  $k$.

The basic latent class model results, if the data are  a collection of multivariate binary observations $\ym_1, \ldots,\ym_N$, where
each $\ym_i=(y_{i1},\ldots,y_{i \ymd})'$   is an $\ymd$-dimensional vector of 0s and 1s, assumed to be the
realization of a binary multivariate random variable $\rvYm=(\rvY_1, \ldots,\rvY_{\ymd})$.
The marginal distribution of $\rvYm$ is then equal to a mixture of
 $\ymd$ independent Bernoulli distributions, with density:
\begin{eqnarray*}
p(\ym_i|\thmod)= \sum_{k=1}^K \eta_k \prod_{j=1}^{\ymd}
\plsw{k}{j} ^{y_{ij}} (1-\plsw{k}{j}) ^{1-y_{ij}},
\label{mix:lcmeq4e}
\end{eqnarray*}
where  $\plsw{k}{j}=\Prob{\rvY_{j}=1|S_i=k}$ is the occurrence probability for each $j=1, \ldots , \ymd$ in the different classes and
the $K$ components of the mixture distribution  correspond to the $K$ latent classes.

Over the years, many  variants and extensions of the basic latent class model have been considered. One
particularly useful extension deals with
 multivariate categorical data  $\ym_1, \ldots,\ym_N$,
where $\ym_i= (y_{i1},\ldots,y_{i \ymd})$ is   the realization of an $\ymd$-dimensional categorical random
variable $\rvYm=(\rvY_1, \ldots,\rvY_{\ymd})$ as above, however, with  each element  $\rvY_j$   taking one value out of
$\catd_j$ categories $\{1,\ldots, \catd_j\}$. Again,   a multivariate mixture distribution  results:
\begin{eqnarray}
p(\ym_i|\thmod)= \sum_{k=1}^K \eta_k \prod_{j=1}^{\ymd}\prod_{l=1}^{\catd_j} \plsw{k}{jl} ^{\indic{y_{ij}=l}}
, \label{mix:lcmcat}
\end{eqnarray}
where $\plsw{k}{jl} = \Prob{\rvY_{j}=l|S_i=k}$ is the probability
of category $l$ for feature $\rvY_{j}$
 in   class  $k$.  Within a Bayesian framework, the $K\ymd$ unknown probability distributions $\plswv{k,j}=(\plsw{k}{j1},\cdots,\plsw{k}{jD_j} )$ of
feature $\rvY_j$ in class $k$ are equipped with a symmetric Dirichlet prior $\plswv{k,j}\sim
\Dirinv{D_j}{g_{0,j}} $. In Step~(a) of Algorithm~\ref{alg1},  this leads to full conditional posterior distributions  $\plswv{k,j}|\Siv, \ym$ arising from
the Dirichlet distribution, see  \citet[Section~9.5]{fru:book} for further details.

If $K$ is unknown,  then  the marginal likelihood  $p(\ym|K)$ could be used to estimate $\hat{p}(\ym|K)$   over a range of different values of $K$,
 using e.g.~bridge sampling \citep{fru:est}.
  A particularly stable  estimator $\hat{p}(\ym|K)$ of the marginal likelihood is given by full permutation  bridge sampling,
  where the importance density is derived from all $K!$ possible permutations $\rho_s$ of  the group labels of a subsequence of posterior draws
$\Siv \im{l}, l=1, \ldots, S_0$ of the unknown allocations, see \citet[Section~7.2.3.2]{cel-etal:mod} for more details.
 Sparse finite as well as DP mixtures of  latent class models are
 interesting alternatives to estimate the number of data clusters in model-based clustering.
 This will be investigated through a simulation study in Section~\ref{sec:sim}.

\subsection{Sparse finite Poisson  mixture models} \label{secpoi}

A popular  model for capturing unobserved heterogeneity and excess zeros in  count data
is the Poisson mixture model, where the data $\ym=(y_1,\ldots,y_N)$
are assumed to be independent  realizations of a random variable $\rvY$ arising from a
finite mixture of Poisson  distributions:
\begin{eqnarray*}
	\rvY  \sim \eta_1 \Poi {\mu_1} + \cdots + \eta_K \Poi{\mu_K}, \label{mix:eye:eq1}
\end{eqnarray*}
with $ \Poi{\mu_k}$ being a Poisson distribution with mean $\mu_k$.  Based on  a Gamma prior,   the full conditional posterior
  $\mu_k|\Siv, \ym$ in  Step~(a) of Algorithm~\ref{alg1} arises  from a Gamma distribution,
 see \citet[Section~9.2]{fru:book} for more details.
An application of a sparse mixture of Poisson  distributions to the  \eye\  \citep{esc-wes:com} will be considered  in Subsection~\ref{seceye}.

To select $K$,  \citet{fru:book} considers  RJMCMC methods, following  \citet{via-etal:bay}, as  well as
marginal likelihoods $p(\ym|K)$. 
Even for this simple mixture with a univariate parameter
$\mu_k$, implementing RJMCMC  required carefully designed split and merge moves.  Concerning marginal likelihoods, bridge sampling with an
importance density obtained from random permutation sampling (see \citet{fru:est}  and \citet[Section~5.4.2]{fru:book}),  turned out to be rather
unstable for  larger values of $K$.  
An alternative estimator $\hat{p}(\ym|K)$ of the marginal likelihood is given by full permutation  bridge sampling,
  where the importance density is derived from all $K!$ possible permutations $\rho_s$ of  the group labels of a subsequence of posterior draws
$\Siv \im{l}, l=1, \ldots, S_0$ of the unknown allocations:
\begin{eqnarray}
q (\mu_1,\ldots,\mu_K,\etav)
= \frac{1}{S_0 K!}
\sum_{l=1}^{S_0}  \sum_{s=1}^{K!} p(\rho_s(\etav)   |\Siv \im{l} ) \prod_{k=1}^K p(\rho_s(\mu_k) | \Siv \im{l}  ,\ym).
\label{QMIX}
\end{eqnarray}
This leads to stable estimators for the marginal likelihood even for larger values of $K$. However,
since the number of functional evaluations increases with $K!$ this method is rather
computer-intensive, and sparse finite Poisson mixture as well as DPM appear to be an attractive alternative.

\subsection{Sparse finite mixtures of GLMs for count data}\label{mixGLM}

Finite mixtures of generalized linear models (GLMs) based on the Poisson, the binomial, the negative binomial, or the multinomial distribution, have found
numerous applications in biology, medicine and  marketing  in order to deal with overdispersion and unobserved heterogeneity; see \citet[Section~9.4]{fru:book} for a review. A finite mixture of Poisson regression models, for instance,  reads:
\begin{eqnarray} \label{mixregpoi}
p(y_i|\theta_1, \ldots, \theta_K,\etav)  = \sum_{k=1}^K \eta_k \Poipdfa{y_i}{\lambda_{k,i}},
\end{eqnarray}
where $\Poipdfa{y_i}{\lambda_{k,i}}$ is the Poisson density with mean  $\lambda_{k,i}=\exp(\xm_i \alpv_k)$,
$\xm_i$ is a row vector containing the observed covariates (including 1 for the intercept) and
$\alpv_1,\ldots,  \alpv_K$ are    unknown component-specific regression parameters.
 A useful extension of (\ref{mixregpoi}) is a model where the Poisson distribution is substituted by a negative binomial distribution with mean
 being equal to $ \lambda_{k,i}$, while allowing at the same time for overdispersion of an unknown degree.
 Sparse finite mixtures of GLMs will be investigated for  the \fault\  \citep{ait:gensc} in Subsection~\ref{secfabric}.


Implementation of Step~(a) in Algorithm~\ref{alg1}   can  be based on any MCMC sampler
   that  delivers draws  from the posterior distribution  $p(\thetav_k |\Siv, \ym)$ of
  a GLM,  with  the outcomes   $y_i$ being restricted  to
   those observations, where  $S_i=k$.
   Various proposals have been
 put forward how to estimate the unknown parameters of  a GLMs for count data
  (including the overdispersion parameter  for negative binomial distributions)  such as
  auxiliary mixture sampling \citep{fru-etal:imp} and the P\'{o}lya-Gamma  sampler \citep{pol-etal:bay_inf}.

  To estimate  $K$ for  a given family of regression models $p(y_i|\thetav_k)$,    marginal likelihoods could be
  computed for each $K$.  This is not at all straightforward for mixtures of GLMs, however  a technique introduced in  \citet{fru-wag:mar}
  can  be used to approximate  the  marginal likelihood $p(\ym|K)$.
 Sparse finite mixtures of GLMs offer an attractive alternative to facing this computational challenge.

\subsection{Sparse finite mixtures  of skew normal and skew-$t$ distributions} \label{mixskew}

 Finally,
 clustering of continuous data with non-Gaussian clusters  using  mixtures of skew normal  and skew-$t$ distributions
 is discussed in this subsection. 
Applications to the univariate  \alz\ \citep{fru-pyn:bay} will be considered in Subsection~\ref{secalz}, whereas  Subsection~\ref{secDLBLC}
considers the multivariate flow cytometric   \DLBCL\ \citep{lee-mcl:mod}.

When clustering continuous data where the clusters are expected to have non-Gaussian shapes,
 it may be  difficult to decide, which (parametric) distribution is appropriate to characterize
the data clusters, especially in higher dimensions.  \citet{mal-etal:ide} pursued a sparse finite mixture of Gaussian mixtures  approach.
 They  exploit the ability of normal mixtures to accurately approximate a
wide class of probability distributions and model the
non-Gaussian cluster distributions themselves by Gaussian mixtures.  On top of that,  they
 use the concept of  sparse finite mixture models  to select the number of  the (semi-parametrically estimated)   non-Gaussian clusters.

On the other hand, many researchers  exploited  mixtures of parametric non-Gaussian
component distributions to cluster such data.
To capture non-Gaussian clusters, many papers consider
skew distributions as introduced by Azzalini  \citep{azz:cla,azz:fur} as component densities,
see e.g.~\citet{fru-pyn:bay} and  \citet{lee-mcl:mod}, among  many others.
 A univariate random variable $X$ follows a
 standard univariate skew normal  distribution with  skewness parameter $\alpha$,  if the pdf takes the  form $p(x)= 2 \phi(x) \Phi(  \alpha x)$,
where $\phi(\cdot)$ and $\Phi(\cdot)$ are, respectively,  the pdf and the cdf of the standard normal distribution.
For $\alpha<0$, a  left-skewed density results,  whereas the density is right-skewed for $\alpha>0$. Obviously, choosing $\alpha=0$
leads back to the  standard normal distribution. The standard skew-$t$ distribution with $\nu$ degrees of freedom results,
if $\phi(\cdot)$ and $\Phi(\cdot)$  are, respectively, the pdf and the  cdf   of  a $t_{\nu}$-distribution.
In a   mixture context,   the skewness parameter $\alpha_k$ and (for  univariate
 skew-$t$ mixtures) the degree of freedom parameter  $\nu_k$   take component-specific values for each mixture component.
  For both families, group-specific location  parameters $\xis_k$ and scale parameters $\omegas_k$ are introduced through the  transformation
  $Y=\xis_k + \omegas_k X$.

  A multivariate version of the skew normal  distribution has been defined in \citet{azz-dal:mul}, while
  multivariate skew-$t$ distributions have been introduced by \citet{azz-cap:dis}. 
  %
 In a multivariate setting, the skewness parameter $\alphav$   is a vector of dimension $r$.
    For standard members of this family,  the pdf  takes the form $p(\bx)= 2 \phi(\bx) \Phi(  \alphav' \bx)$  with
     $\phi(\cdot)$ and  $\Phi(\cdot)$   being  equal to, respectively,  the pdf  of   the $r$-variate  
     and   the cdf   of the univariate  standard normal distribution for   the  multivariate skew normal distribution.
   For the  multivariate skew-$t$  distribution with $\nu$ degrees of freedom,
    $\phi(\cdot)$ and  $\Phi(\cdot)$   are equal to, respectively,  the pdf  of  the $r$-variate  
    and  the cdf   of the univariate
    $t_{\nu}$-distribution.
   As for the univariate case, 
   group-specific location  parameters
   $\xiv_k$ (a vector of dimension $r$) and scale matrices $\Omegas_k$ (a matrix of dimension $r \times r$) are introduced through the  transformation
  $\rvYm=\xiv_k + \Omegas_k  \rvXm$, where $\rvXm$ follows the standard $r$-variate distribution described above, with  component-specific skewness
  parameters $\alphav_k$ and (for  multivariate  skew-$t$ mixtures)   component-specific degrees of freedom parameters  $\nu_k$.
%
%

The first paper which  considered Bayesian  inference, both for univariate as well as multivariate mixtures of skew normal and skew-$t$ distributions,
is  \citet{fru-pyn:bay} who developed an efficient MCMC scheme, combining a latent variable  representation with a latent factor following a truncated
 standard normal distribution with data augmentation. This MCMC scheme can be easily  incorporated in Step~(a) of Algorithm~\ref{alg1}
  to estimate sparse finite mixtures  of skew normal and skew-$t$ distributions as well as DPM.
 \citet{fru-pyn:bay} also  discussed  various  methods for selecting $K$ for  finite mixtures
of skew normal and skew-$t$ distributions, both in the univariate as well as in the multivariate case,
among them marginal likelihoods $p(\ym| K)$ computed using bridge sampling \citep{fru:est},
BIC and various  DIC criteria   \citep{cel-etal:dev}. 
However, it was practically
impossible to compute the marginal likelihood $p(\ym| K)$ for mixtures with more than 5 or 6 components.
Hence, sparse finite mixtures of skew normal and skew-$t$  distributions appear to be an attractive
way to select the number of groups or clusters for such mixture models.

\section{A Simulation Study}  \label{sec:sim}

\begin{table}[t]
	\centering
	\begin{tabular}{|c|ccc|ccc|cccc|}
		\hline
		& \multicolumn{3}{c|}{$Y_1$} &\multicolumn{3}{c|}{$Y_2$}& \multicolumn{4}{c|}{$Y_3$} \\
		Categories	& 1& 2&3& 1& 2& 3& 1& 2&3 &4\\	\hline
		Class 1		& 0.1 & 0.1&0.8  &0.1& 0.7& 0.2& 0.7& 0.1& 0.1& 0.1 \\
		Class 2		& 0.2 & 0.6&0.2  &0.2& 0.2& 0.6& 0.2& 0.1& 0.1& 0.6 \\
		\hline
	\end{tabular}
	\caption{Occurrence probabilities for the three variables in the two classes. } \label{tab1}
\end{table}

 The aim of this simulation study is to investigate whether 1) a sparse finite mixture of non-Gaussian components   appropriately estimates the number of data clusters,
	2) the posterior of $K_+$  of sparse finite mixtures and DPM is comparable, if the priors on the precision parameters $e_0$ and   $\alpha$  are matched,
 and 3) whether both approaches estimate similar partitions of the data.
 Additionally, the impact of the prior on $\alpha$ and $e_0$, the number of specified components $K$,   and  the number of observations $N$ is investigated.

Inspired by the \fear\   which will be analyzed in Subsection~\ref{secfear},  we  generate  multivariate categorical data using
following simulation setup.
100 data sets with, respectively,   $N=100$ and  $N=1000$  observations   are simulated from a latent class model  with two classes of  equal size (i.e.~$\eta_1=\eta_2=0.5$) and three  variables with
$D_1=3$, $D_2= 3$, and $D_3=4$ categories. The occurrence probabilities are given in Table~\ref{tab1}.
 Sparse latent class models  with $K=10$ and  $K=20$ as well as  DPM are fitted to each data set.
    For both model classes,  the Gibbs sampler  is run using Algorithm~\ref{alg1} for 8000 iterations after discarding 8000 draws as burn-in.
 The starting classification  is obtained by clustering the data points into  $K=10$ or $K=20$ clusters using $k$-means.

Various  priors $\alphaDP \sim \Gammad{\ala ,\alb}$ on the precision parameter $\alphaDP$  of the DPM are investigated 
 and  matched to the prior $\ed{0} \sim \Gammad{\ala ,K \alb}$ on the precision parameter $ \ed{0}$  of the sparse latent class model
 as described in Subsection~\ref{bay_hyp}.  The first prior,  $\alpha \sim\cG(1,20)$   with $\Ew{\alphaDP} =  0.05$, corresponds to
  the sparse priors $e_0\sim \cG(1,200)$ (for $K=10$) and  $e_0\sim \cG(1,400)$  (for $K=20$) and yields a \lq\lq sparse\rq\rq\ DPM. The remaining two priors, $\alphaDP \sim\cG(1,2)$ and $\alphaDP \sim\cG(2,1)$,  with  $\Ew{\alphaDP} =  0.5$  and $2$  reflect common choices in the literature.

\begin{table}[ht]
	\centering
	\begin{small}  
		\begin{tabular}{llcccccccc}
			\hline	
			prior&method&   &$K_+=1$& $K_+=2$ & $K_+=3$ & $K_+=4$ & $K_+=5$	& $K_+=6$ & $K_+\ge7$  \\ %
			\hline
			$\alpha\sim\cG(1,20)$
			&SFM& $K=10$&0.000 &\textbf{0.813}& 0.166  & 0.019 
			&0.002   &  0.000 & 0.000     \\
			&& $K=20$&0.000 &\textbf{0.812} & 0.162  & 0.022 
			&0.003      & 0.001 & 0.000     \\
			&DPM&&0.000		 &\textbf{ 0.704} & 0.252& 0.040  
			&0.004	              &  0.000  & 0.000  \\
		\hline
			$\alpha\sim\cG(1,2)$
			&SFM&$K=10$&0.000 & 0.310 	 & \textbf{0.367}  & 0.210 
			&0.082 &0.025  & 0.006   \\
			&  &$K=20$&0.000 &\textbf{ 0.359} 	 & 0.320  & 0.178  
			& 0.085 & 0.035  & 0.023    \\
			&DPM&&0.000		 & \textbf{ 0.345}  & 0.312& 0.199  
			& 0.095		 & 0.035 & 0.015 \\	
		\hline
			$\alpha\sim\cG(2,1)$
			&SFM&$K=10$&0.000 & 0.094 	 & 0.207  &\textbf{ 0.237 }  
			&0.200 & 0.140  & 0.124   \\
			&   &$K=20$&0.003 & 0.123  & 0.188  & \textbf{0.210}    
			&0.179 & 0.135  & 0.158  \\
			&DPM&&0.000 & 0.099  & 0.188&\textbf{ 0.210}  
			&0.188		 & 0.133  & 0.174\\
			\hline
		\end{tabular}
		\caption{Posterior distribution $p(K_+|\by)$ for various prior specifications on $e_0$ and $\alpha$,
 for $K=10$ and $K=20$, for the first data set of the simulation study, $N=100$.} \label{tab2}
	\end{small} 
\end{table}

\begin{table}[ht]
	\centering
	\begin{small}  
      \begin{tabular}{llccccccccc}
			\hline
			& & &\multicolumn{4}{c}{$N=100$} & \multicolumn{4}{c}{$N=1000$} \\	
			prior&method&   &$\Ew{p.p.|\ym}$		& $\Knhat$ & ari  & err & $\Ew{p.p.|\ym}$	 	& $\Knhat$ & ari  & err\\ 
			\hline
			$\alpha\sim\cG(1,20)$
			&SFM& $K=10$&0.009 &	1.94& 0.44  & 0.18 
                                            &0.010   &  2.05 & 0.54  & 0.13    \\
			&& $K=20$&0.005 &	1.92 & 0.43  & 0.18 
                                                &0.005      & 2.02 & 0.54  & 0.13    \\
			&DPM&&0.092		 & 1.99 & 0.44& 0.18  
                                                 &0.110	              &  2.29  & 0.53& 0.14  \\
			\hline
			$\alpha\sim\cG(1,2)$
		&SFM&$K=10$&0.064 & 2.29 	 & 0.46  & 0.17 
                                    &0.068 &2.23  & 0.53  & 0.14 \\
			&  &$K=20$&0.035 & 2.38 	 & 0.45  & 0.17  
                               & 0.032 & 2.24  & 0.53  & 0.14   \\
			&DPM&&0.599		 &  2.44  & 0.45& 0.17  
                            & 0.670		 & 2.62 & 0.52& 0.15 \\	
			\hline
			$\alpha\sim\cG(2,1)$
			&SFM&$K=10$&0.189 & 3.56 	 & 0.45  & 0.19   
                                  &0.163 & 2.97  & 0.52  & 0.15 \\
			&   &$K=20$&0.086 & 3.34  & 0.45  & 0.19    
                              &0.072 & 3.28  & 0.51  & 0.16\\
			&DPM&&1.517		 &  3.50  & 0.44& 0.19  
                                 &1.360		 & 3.72  & 0.49& 0.17\\
			\hline \hline				
			&poLCA	&	 			&& 1.37 & 0.18& 0.35 && 2.00 & 0.54& 0.13 \\
			\hline
		\end{tabular}
		\caption{Average clustering results  over 100 data sets of size $N=100$ and $N=1000$, simulated from a latent class model with two classes, obtained through sparse latent class models (SFM) with $K=10$ and $K=20$ and DPM  for three different priors on the precision parameters $\ed{0}$ and $\alphaDP$
as well as  using EM estimation as implemented in the R package poLCA \citep{Mix:LinzerLewis2011}.  The  reported values are averages  of  the posterior expectation $\Ew{p.p.|\ym}$  of the precision parameter $\ed{0}$ (SFM) and $\alphaDP$ (DPM), the estimated number of clusters  $\Knhat$, the  adjusted Rand index (ari)  and the error rate (err).} \label{tab3}
	\end{small} 
\end{table}

The posterior distributions of $K_+$  under the various prior settings are exemplified for one data set in Table~\ref{tab2}. They look similar for DPM
and sparse finite mixture models if the priors are matched accordingly.  The average clustering results  over all data sets, for both $N=100$  and $N=1000$, are reported in  Table~\ref{tab3}. 
The  cluster quality of all estimated  partitions is measured using  the adjusted  Rand index (ari)  \citep{Mix:HubertArabie1985} and the error rate (err)
which is calculated as the proportion of misclassified data points.
For $N=100$, again  the clustering results are  very similar for  DPM and sparse finite mixtures, regardless whether $ K=10$ or $K=20$,  or smaller or larger expected values for  $e_0$ and $\alphaDP$ are  defined, as long as the hyper priors  are matched. For the sparse hyper priors  $\alphaDP \sim \cG(1,20)$ and $e_0 \sim \cG(1,20K)$,  the average  of the posterior mode estimators $\Knhat$ over all  data sets is very close to $2$,
whereas for more common priors  on  $\alphaDP$ 
 this  average is  considerably larger than 2,  both  for sparse latent class models and DPM. However, the adjusted Rand index   and the error rate  are roughly the same for all priors, indicating that the superfluous clusters only consist  of a few observations.
The results for larger  data sets with $N=1000$ observations lead to similar  conclusions, with the   DPM  showing  a stronger  tendency toward overfitting  $\Knhat$ than sparse finite mixtures, despite  matching the hyper priors for the precision parameters.





	For comparison, for each data set  a standard latent class analysis is performed using the EM algorithm and the BIC  criterion to estimate the number of clusters.  The R package poLCA \citep{Mix:LinzerLewis2011} is used for this  estimation.
For $N=100$, the poLCA approach underestimates the number of data clusters, probably  because the asymptotic consistency  of   BIC does not apply to small-sized data sets.  For $N=1000$,  the poLCA approach performs equally well as the  sparse finite mixture approach.

The simulation study also provides evidence that specifying  a (sparse) hyper prior over  $e_0$ is preferable to
 choosing a fixed (small) value.  As shown in Figure~\ref{plot:priorKplus}  for $N=100$, a sparse finite mixture
with $K=10$ and fixed value $e_0=0.005$  basically prefers  a one-cluster solution.
However, as can be seen from the first row in Table~\ref{tab3}, by specifying the
prior $e_0\sim \cG(1,200)$   the posterior mean $\Ew{e_0|\ym}$   is on average twice as large as the  prior mean $\Ew{e_0}= 0.005$ and
on average 1.94 clusters are estimated, meaning that one cluster was selected for \g{only  few data sets. }

\section{Applications}  \label{sec:appl}


For each type of   mixture models discussed in Section \ref{section4},
a case study is  provided  to compare  sparse finite mixtures with  DPM of the same type.
For both model classes, the influence of the  priors
$p( \ed{0})$ and $p(\alphaDP)$  on the posterior distribution   $p( \Kn|\ym)$   of the number of clusters  $\Kn$ is investigated  in detail.
 Typically,  for sparse finite mixtures $K=10$ and   $\ed{0} \sim \Gammad{1 ,200}$,
implying    $\Ew{\ed{0}}=0.005$,  is specified whereas  for DPM 
 $\alphaDP \sim\cG(2,4)$ is specified  as in \citet{esc-wes:bay}.  In addition,   both priors  are matched  as described in Subsection~\ref{bay_hyp}.
%
 For each case study,    standard finite mixtures  with $ \ed{0}=4$  are estimated  for increasing $K$.

\subsection{Application to the \fear\ }  \label{secfear}

  \citet{ste-etal:sta} consider data of  $N=93$ children from white middle class homes in the U.S.,  tested at age 4 and 14 months,
in the context of infant temperamental research.
For each child, three  categorical data (i.e.~multivariate data of  dimension $\ymd=3$)  are observed, namely
 motor activity (M) at 4 months with $\catd_1=4$ categories,
 fret/cry behavior (C) at 4 months  with $\catd_2=3$   categories,
and  fear of unfamiliar events (F) at 14 months  with $\catd_3=3$   categories, see Table~\ref{feradta}. The categories can be interpreted as scores with higher scores indicating a stronger behavior.

\begin{table}
\begin{center}
\begin{tabular}{cc}
\begin{tabular}{|l|l|ccc|}
  \hline
  & & F=1 & F=2 & F=3 \\
  \hline
M=1 &  C=1 &   5   &  4  &   1 \\
  & C=2 &  0   &  1  &   2\\
  & C=3 &    2   &  0  &   2\\
  \hline
  M=2 &  C=1 &   15   &  4 &    2\\
   & C=2 &  2  &   3  &   1\\
   & C=3 &    4 &    4 &    2\\
     \hline
\end{tabular}&
\begin{tabular}{|l|l|ccc|}
\hline
 & & F=1 & F=2 & F=3 \\
 \hline
  M=3 &  C=1 &     3 &    3  &   4\\
  & C=2 &   0  &   2   &  3\\
  & C=3 &    1  &   1   &  7\\
\hline
 M=4 &  C=1 &     2   &  1 &    2\\
  & C=2 &   0  &   1  &   3\\
  & C=3 &    0  &   3  &   3\\
\hline
\end{tabular}
\end{tabular}
\caption{\fear;  $4 \times 3 \times 3$ contingency table summarizing the data   which measure motor activity (M) at 4 months,
    fret/cry behavior (C) at 4 months,  and  fear of unfamiliar events (F) at 14 months for $N=93$  children  \citep{ste-etal:sta}.} \label{feradta}
\end{center}
\end{table}

The scientific hypothesis is that two different profiles in children are present, called inhibited and unhibited to the unfamiliar
(i.e. avoidance or approach to unfamiliar children, situations and objects). To test this hypothesis, a  latent class model as in (\ref{mix:lcmcat}) is applied,
\begin{eqnarray*}
\Prob{M=m,C=c,F=f}  = \sum_{{  k}=1}^K \eta_{ k}  \pi^M_{{ k},m}  \pi^C_{{  k},c} \pi^F_ {{ k},f}, \label{mix:nlterg}
\end{eqnarray*}
with class specific probability distributions
 $\piv^M_{{  k}}=(\pi^M_{{  k},1}, \ldots, \pi^M_{{  k},4})$,
 $\piv^C_{{  k}}=(\pi^C_{{  k},1}, \ldots, \pi^C_{{ k},3})$, and
 $\piv^F_{{  k}}=(\pi^F_{{ k},1}, \ldots, \pi^F_{{ k},3})$ and $K$ being unknown.

Three types of mixture models are considered,  assuming the class specific probability distributions   $\piv^M_{{  k}}$,
 $\piv^C_{{  k}}$, and  $\piv^F_{{  k}}$   to be independent, each following
    a symmetric Dirichlet prior  $\Dirinv{D_j}{g_{0,j}}$   
    with $g_{0,j}=1$ for $j=1, \ldots,3$.
Sparse latent class models   as described in Subsection \ref{seclcm} are estimated with $K=10$ and compared to
    DP latent class models.  In addition,  a  standard latent class model  with $ \ed{0}=4$ is
estimated for increasing $K$  and   marginal likelihoods are
  computed using full permutation bridge sampling, see Table~\ref{tab:fear-bic}.

\begin{figure}[h]
\begin{center}
\scalebox{0.4}{\includegraphics{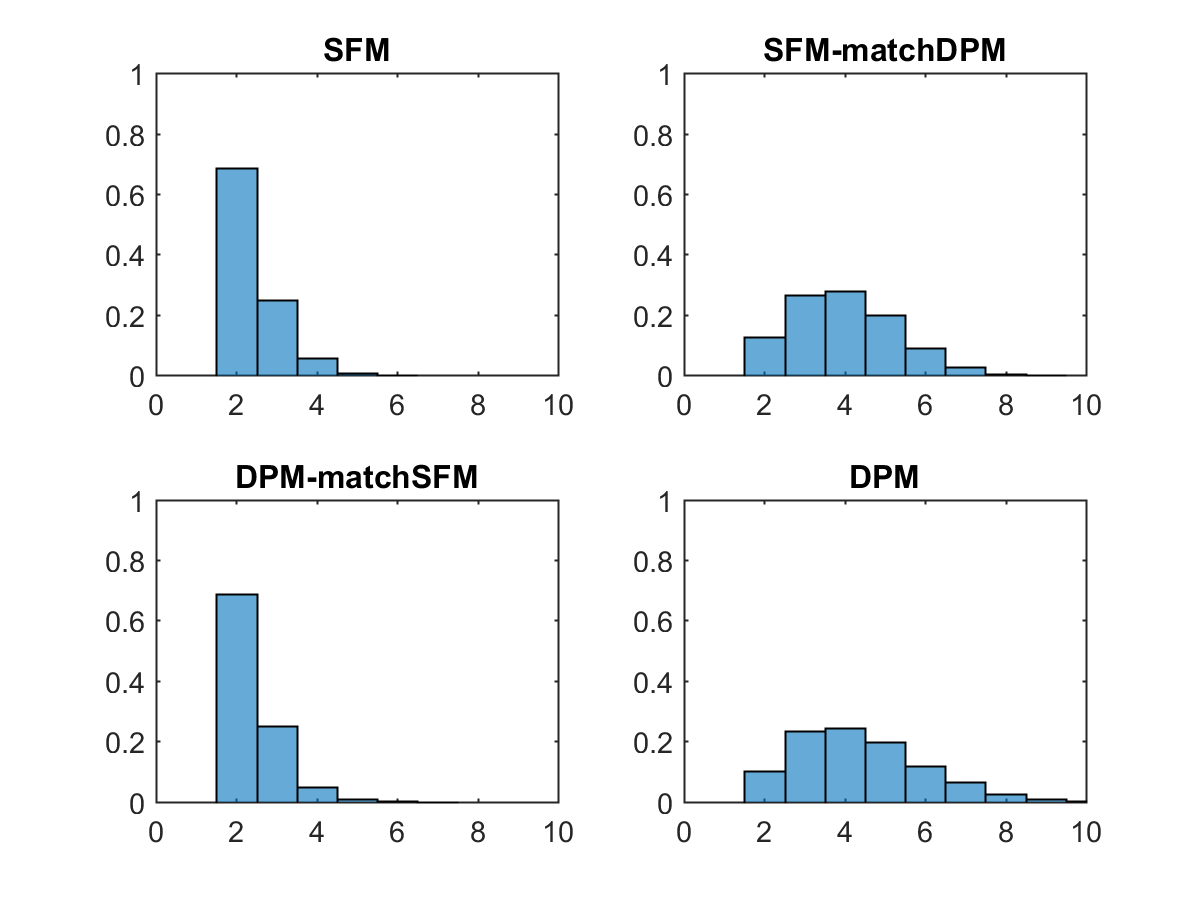}}
 \end{center}
 \caption{\fear ;  posterior distributions  $\Prob{\Kn |\ym}$ of the number of clusters $ \Kn$;
  top:  sparse finite mixtures  with  $K=10$,   $e_0\sim \cG(1,200)$ (left hand side) and  matched prior $e_0\sim \cG(2,4 K)$ (right hand side);
 bottom: DPM with $\alphaDP \sim \cG(2,4)$ (right hand side) and  matched prior  $\alphaDP \sim \cG(1,200/K)$ (left hand side).}\label{fear_postK}
 \end{figure}

Table~\ref{tab:fear-bic} and Figure~\ref{fear_postK}  
compare the various   posterior distributions  $\Prob{\Kn |\ym}$ of the number of clusters $ \Kn$ under the specific hyper priors.
  Both for the marginal likelihood as well as for a sparse finite mixture, $\Knhat=2$ is selected, confirming  the theoretically expected
number of clusters,  whereas the DPM overestimates the number of clusters
   with $\Knhat=4$.
   However, once the hyper prior for $\alphaDP$ is  matched  to the sparse finite mixture,  the resulting \lq\lq sparse\rq\rq\ DPM also selects
   two clusters. On the other hand,   a   sparse  finite mixture matched to the DPM is overfitting. 
   This  example illustrates the importance of  prior shrinkage  of $\ed{0}$  and $\alphaDP$  towards small values.

In Table~\ref{tab_fear_K2}, the estimated occurrence probabilities for the two classes are reported. Clearly,  the children in the two classes have a rather different profile. Whereas children belonging to class 1 are more likely to have higher scores in all three variables, children in class 2 show less motor activity, crying behavior and fear at the same time. This clustering result is in line with the psychological theory behind the experiments, according to which all three behavioral variables are regularized by the same physiological mechanism, see \cite{ste-etal:sta} for more details.

\begin{table}
{\small
\begin{center}
\begin{tabular}{lccccccc}
\hline
$\Prob{\Kn  |\ym}$    & $\Kn=1$ & $\Kn=2$ &$\Kn=3$ &$\Kn=4$ &$\Kn=5$ &$\Kn=6$ &$\Kn\geq7$  \\
   \hline
   SFM 
   & & & & & & & \\
  \hspace*{3mm}  $  \ed{0} \sim \Gammad{1 ,200}$  &    0 &  \textbf{0.686} &  0.249 &  0.058 &  0.007 &  0.001 & 0.000 \\
 \hspace*{3mm}  matched to   DPM &0 &  0.128 &  0.267  &  \textbf{0.280} &  0.201 &  0.090 & 0.033  \\ 
  \hline
 DPM  & & & & & & & \\
 \hspace*{3mm} $\alphaDP \sim \Gammad{2,4}$  &0 &  0.101 &  0.235 &  \textbf{0.246 }&  0.197 &  0.118 & 0.103 \\ 
\hspace*{3mm}   matched to SFM                 & 0 &  \textbf{0.688} &  0.251 &  0.048 &  0.011 &  0.002 &  0.000 \\
  \hline
 \end{tabular}\\[1mm]
 \begin{tabular}{lccccc}
\hline
$\log \hat{p} (\ym|K)$  & $K=1$ & $K=2$ &$K=3$ &$K=4$ &$K=5$  \\
 \hline
%
  FM ($e_0=4$)   & -333.01 & {\bf -330.46} &  -333.67 &  -337.37 &  -340.48  \\
  \hline
\end{tabular}
\caption{\fear ;  the rows in the upper table  show  the posterior distribution  $\Prob{\Kn |\ym}$ of the number of clusters $ \Kn$ for
various latent class models:  sparse latent class models  with  $K=10$ (SFM) with  hyper priors $e_0\sim \cG(1,200)$ and $e_0\sim \cG(2,4 K)$  (matched to DPM),
   DPM with hyper priors $\alphaDP \sim \cG(2,4)$  and $\alphaDP \sim \cG(1,200/K)$ (matched to SFM). The lower  table shows
  log marginal likelihoods, $\log \hat{p}  (\ym|K)$,  estimated for a  latent class model  with $e_0=4$  (FM) for  increasing  $K$.}\label{tab:fear-bic}
\end{center}}
\end{table}

\begin{table}
\begin{center}
\begin{tabular}{lccc}
      \hline
   &  Class 1 &  Class 2\\ \hline
     $\pi^M_{k,1}$  &   0.146 (0.032,0.267) &   0.225 (0.103,0.358)\\
    $\pi^M_{k,2}$ & 0.170 (0.010,0.319)  & {\bf 0.573} (0.408,0.730)\\
 $\pi^M_{k,3}$ &   {\bf 0.408} (0.243,0.578) &  0.126 (0.015,0.239)\\
  $\pi^M_{k,4}$ &   0.276 (0.127,0.418)  &  0.076 (0.002,0.159)\\
    \hline
$\pi^C_{k,1}$  &  0.263 (0.078,0.419) & {\bf 0.679} (0.519,0.844) \\
$\pi^C_{k,2}$ &   0.311 (0.170,0.478) &  0.109 (0.007,0.212)\\
$\pi^C_{k,3}$ &  {\bf 0.426}  (0.261,0.598) &  0.212 (0.079,0.348)\\
    \hline
$\pi^F_{k,1}$  &   0.069 (0.000,0.177) &  {\bf 0.629} (0.441,0.823)\\
  $\pi^F_{k,2}$  &  0.298 (0.119,0.480)  &  0.279 (0.117,0.447)\\
  $\pi^F_{k,3}$  & {\bf 0.633} (0.447,0.830) &  0.090 (0.000,0.211)\\
\hline
   $\eta_k$   & 0.470 (0.303,0.645) &  0.530 (0.355,0.698)\\
    \hline
\end{tabular}
\caption{\fear ; posterior inference for $\piv^M_{{  k}}$,
 $\piv^C_{{  k}}$, and  $\piv^F_{{  k}}$, based on all MCMC draws with  $\Kn = 2$. The values are the average of the MCMC draws,
 with  95\% HPD intervals in parentheses.}  \label{tab_fear_K2}
  \end{center}
\end{table}


\begin{table}
{ \small
\begin{center}
\begin{tabular}{lccccccc}
\hline
 $\Prob{\Kn  |\ym}$  & $\Kn=1,2$ &$\Kn=3$ &$\Kn=4$ &$\Kn=5$ &$\Kn=6$ &$\Kn=7$  & $\Kn\geq8$ \\
   \hline
   SFM  
   & & & & & & & \\
  \hspace*{3mm}  $  \ed{0} \sim \Gammad{1 ,200}$  &   0.000  &  0.091 &  {\bf 0.584}  &  0.266 &  0.056 &  0.003 & 0.000 \\
 \hspace*{3mm}  matched to   DPM                            & 0.000  &  0.007 &  0.174 &  {\bf 0.308} &  0.299 &  0.153 &  0.059 \\ 
\hline
 DPM  & & & & & & & \\
 \hspace*{3mm} $\alphaDP \sim \Gammad{2,4}$   &  0.005 &  0.095 &  0.209 & {\bf  0.222} &  0.173 &  0.134 &  0.161 \\
\hspace*{3mm}   matched to SFM  & 0.000  &  0.012 &  {\bf  0.464} &  0.379 &  0.122 &  0.022 &  0.002 \\
  \hline
  \end{tabular}\\[1mm]
  \begin{tabular}{lccccccc}    \hline
$\log \hat{p} (\ym|K)$  & $K=1$ & $K=2$ &$K=3$ &$K=4$ &$K=5$ &$K=6$ &$K=7$ \\
 \hline
%
  FM  ($e_0=4$)  &
  -472.89 &  -254.19 &  -239.79 &  -234.48 &  -232.9 &  -231.84 &  -231.04 \\
  \hline
\end{tabular}
\caption{\eye ;  the rows in the upper table  show  the posterior distribution  $\Prob{\Kn |\ym}$ of the number of clusters $ \Kn$ for
following Poisson mixture models:  sparse finite  mixtures  with  $K=10$ (SFM) with  hyper priors  $e_0\sim \cG(1,200)$ and $e_0\sim \cG(2,4 K)$
 (matched to DPM),
   DPM  with hyper priors $\alphaDP \sim \cG(2,4)$  and $\alphaDP \sim \cG(1,200/K)$ (matched to SFM). The lower  table shows
  log marginal likelihoods, $\log \hat{p}  (\ym|K)$,  estimated for a  finite mixture  with $e_0=4$ (FM) for  increasing  $K$.}
\label{tab:eye-bic}
\end{center}
}
\end{table}

\clearpage

\subsection{Application to the \eye\ }  \label{seceye}
%
The count data on eye tracking anomalies in 101 schizo\-phre\-nic patients studied
by \citet{esc-wes:com} are reconsidered.  To capture overdispersion and excess zeros diagnosed for this data set,
  \citet{fru:book} analyzed  the data by a  finite Poisson mixture model.
The goal of the analysis is not primarily  clustering of the data, but  capturing the extreme unobserved heterogeneity present in this data set,
using both sparse finite Poisson mixtures with $K=10$ as in Subsection~\ref{secpoi} as well as  DPM.

\begin{figure}[h]
\begin{center}
\scalebox{0.4}{\includegraphics{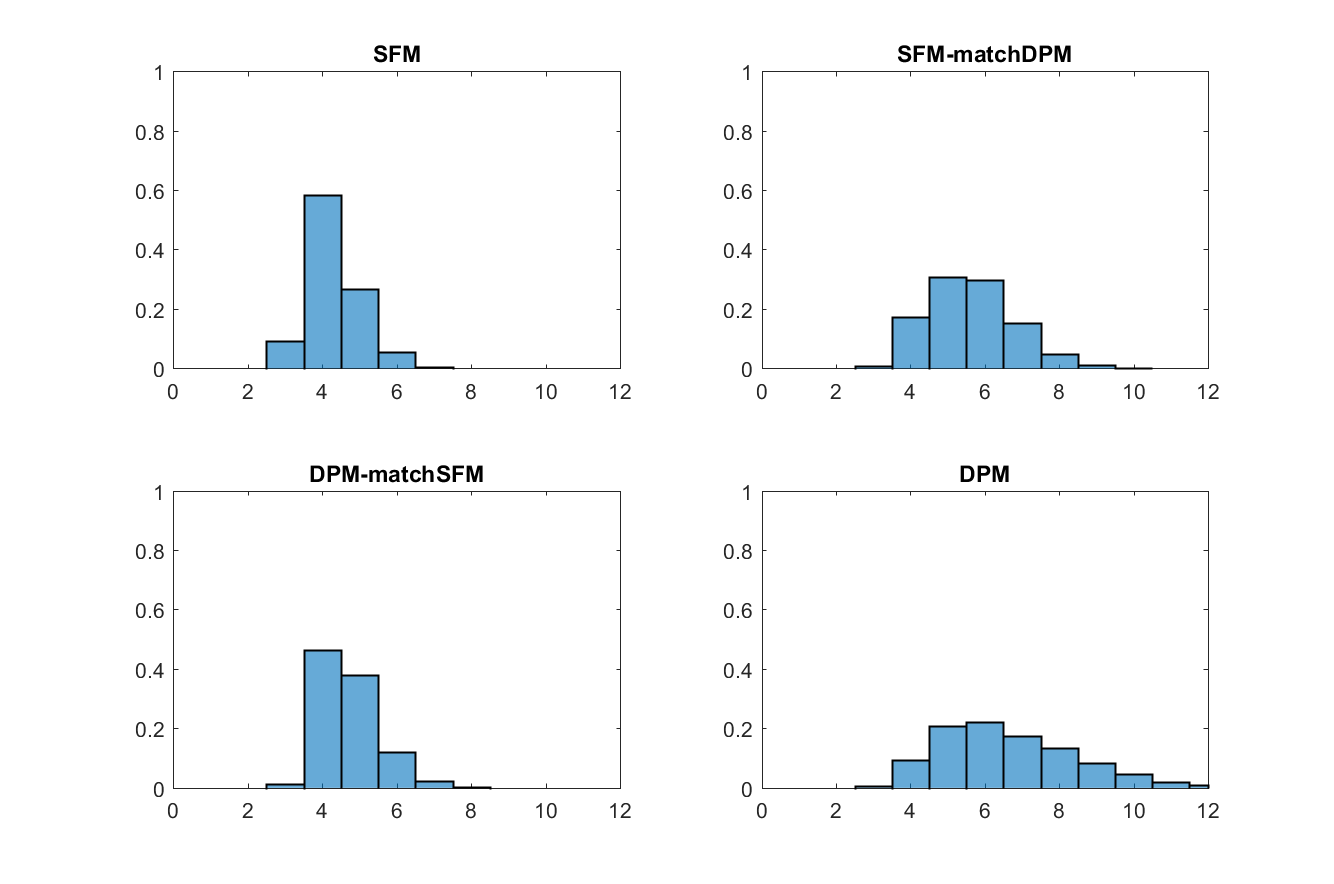}}
 \end{center}
 \caption{\eye ;  posterior distributions  $\Prob{\Kn |\ym}$ of the number of clusters $ \Kn$;   top:  sparse finite mixtures  with  $K=10$,   $e_0\sim \cG(1,200)$ (left hand side) and  matched prior $e_0\sim \cG(2,4 K)$ (right hand side);  bottom: DPM with $\alphaDP \sim \cG(2,4)$ (right hand side) and  matched prior  $\alphaDP \sim \cG(1,200/K)$ (left hand side).}\label{eye_postK}
 \end{figure}

   For all types of mixture models, the same hierarchical prior is applied for the component-specific parameters
 with  $\mu_k|b_0 \sim \Gammad{a_{0},b_{0}}$ and $ b_0 \sim \Gammad{g_0,G_0}$,
   where $a_0=0.1$, $g_0=0.5$, and $G_0=g_0 \mean{y}/a_0$,  with $\mean{y}$ being the mean of the data.
%
 Table~\ref{tab:eye-bic} and Figure~\ref{eye_postK}  
compare the various   posterior distributions  $\Prob{\Kn |\ym}$ of the number of clusters $ \Kn$ under various hyper priors.
The sparse finite Poisson mixture model clearly identifies  four  clusters, whereas the posterior $\Prob{\Kn |\ym}$ is much more spread out for the corresponding DPM, reflecting the extreme unobserved heterogeneity  in the observed counts.
However, once the hyper prior for $\alphaDP$ is  matched  to the sparse finite mixture,  the resulting  DPM also selects
   four clusters. On the other hand,   a   sparse  finite mixture matched to the DPM also indicates considerable unobserved heterogeneity which is confirmed by the marginal likelihoods which are computed using full permutation bridge sampling.


\subsection{Application to the \fault\ }  \label{secfabric}
   For further illustration, we  consider regression analysis of  (count) data on fabric faults \citep{ait:gensc} where
 the response variable $y_i$ is the number of faults in a bolt of length $l_i$.
The goal of the analysis is testing homogeneity, i.e. to investigate if  a single count data regression model is appropriate or whether unobserved heterogeneity is present.
 Based on the regressor matrix $\xm_i =\left(1 \,\ \log l_i\right)$,   mixtures of Poisson and negative binomial regression models are fitted as described in Subsection~\ref{mixGLM}.
 %
 Marginal likelihoods for these data were  computed in  \citet{fru-etal:imp}  
  for standard finite mixture models  with $ \ed{0}=4$ up to $K=4$ and are compared with
   sparse finite GLMs with $K=10$ and  DPM of GLMs in  Table~\ref{tab:fault-bic}.
 For all  mixtures, a priori the component-specific regression coefficients are assumed to be \iid\  from
   a  $\Normal{0,4}$-distribution.     For the negative binomial distribution, the same prior as in    \citet{fru-etal:imp} is assumed for
    the   group specific  degrees of freedom parameter
   $\rho_k$:  $p(\rho_k) \propto
   2d\rho_k/(\rho_k+c)^3$, where the choice of  $c=10/(1+\sqrt{2})$ implies a  prior  median of  10.

\begin{table}
{ \small
\begin{center}
\begin{tabular}{c}
\begin{tabular}{lllcccc}
\hline
$\Prob{\Kn  |\ym}$ & &      & $\Kn=1$ & $\Kn=2$ &$\Kn=3$ &$\Kn=4$   \\
   \hline
Poisson GLM   &   SFM  
 &  $  \ed{0} \sim \Gammad{1 ,200}$  &  0.241 &  \textbf{0.754} &  0.006 & 0.000 \\
                                             &  &  matched to   DPM & 0.060  & \textbf{ 0.887} &  0.053 &  0.001  \\\cline{2-7}
& DPM   &$\alphaDP \sim \Gammad{2,4}$  & 0.036 &  \textbf{0.914} &  0.049 &  0.001  \\
&  &  matched to SFM  &0.141 & \textbf{ 0.832} &  0.027 &  0.000  \\
 \hline
NegBin  GLM &  SFM  
&   $  \ed{0} \sim \Gammad{1 ,200}$  &  \textbf{ 0.994} &  0.006 & &   \\
&  & matched to   DPM                   &\textbf{ 0.906} &  0.093 &  0.001 &   \\\cline{2-7}
& DPM   & $\alphaDP \sim \Gammad{2,4}$  &  \textbf{0.940}&  0.059 &  0.001 & \\
&  &  matched to SFM                        & \textbf{0.994} &  0.006 &  &  \\
  \hline
  \end{tabular}
  \vspace*{2mm} \\
   \begin{tabular}{lccccc}
   \hline
  $\log \hat{p}(\ym|K)$ & & $K=1$ & $K=2$ &$K=3$ &$K=4$ \\
 \hline
Poisson GLM  & FM ($e_0=4$)  &  -101.79 & {\bf -99.21}& -100.74 & -103.21 \\
NegBin GLM  & FM ($e_0=4$)  &  {\bf  -96.04} & -99.05 &  -102.61 &  -105.7 \\
\hline
 \end{tabular}
  \end{tabular}
\caption{\fault ;  the rows in the upper table  show  the posterior distribution  $\Prob{\Kn |\ym}$ of the number of clusters $ \Kn$ for
following mixtures   of Poisson GLMs  and negative binomial GLMs: sparse finite mixtures
 with  $K=10$ (SFM) with  hyper priors  $e_0\sim \cG(1,200)$ and $e_0\sim \cG(2,4 K)$
 (matched to DPM),
   DPM  with hyper priors $\alphaDP \sim \cG(2,4)$  and $\alphaDP \sim \cG(1,200/K)$ (matched to SFM). The lower  table shows
  log marginal likelihoods, $\log \hat{p}  (\ym|K)$,  estimated for    finite  mixtures with $e_0=4$ (FM)  for  increasing  $K$.}
\label{tab:fault-bic}
\end{center}}\end{table}


\begin{figure}[h]\begin{center}
\begin{tabular}{cc}
\scalebox{0.3}{\includegraphics{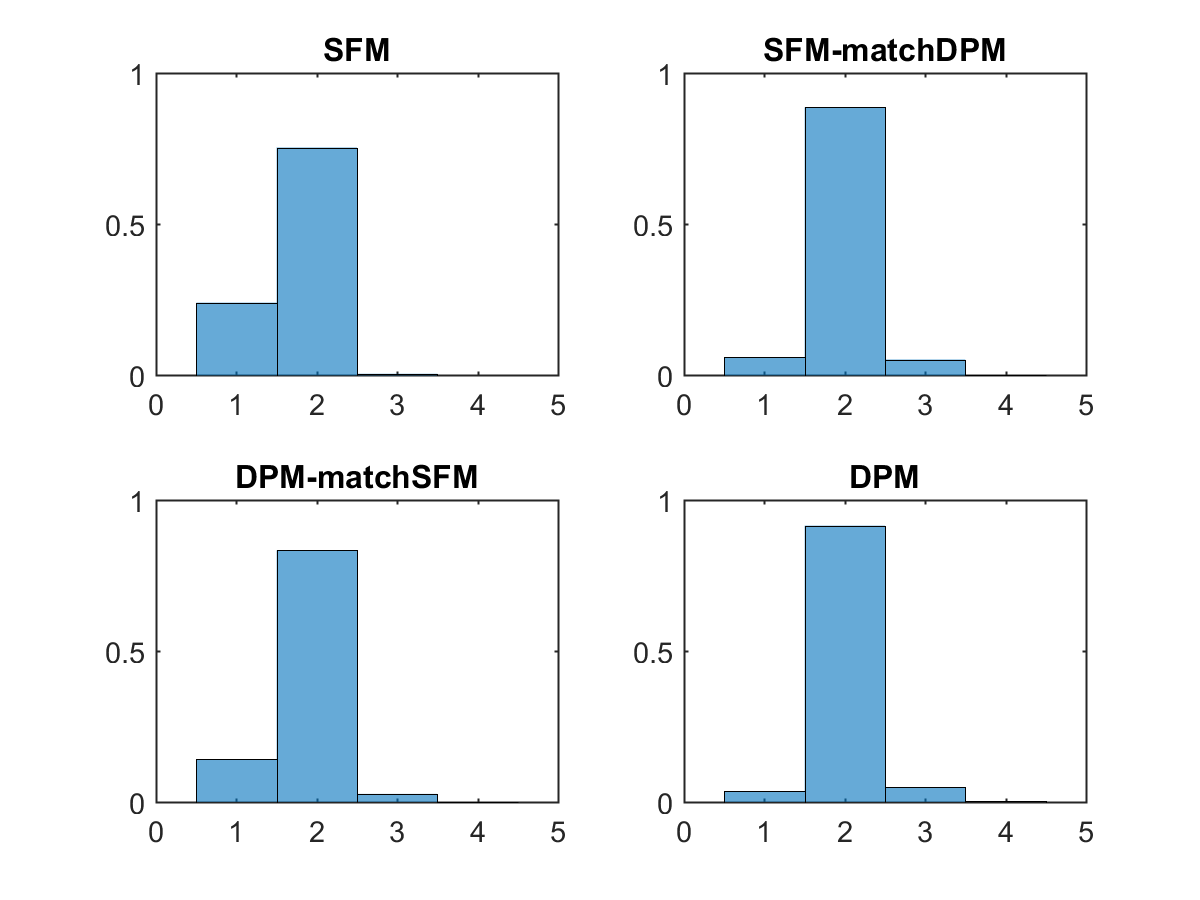}}  \hspace*{0.5cm}  &  \hspace*{0.5cm}
 \scalebox{0.3}{\includegraphics{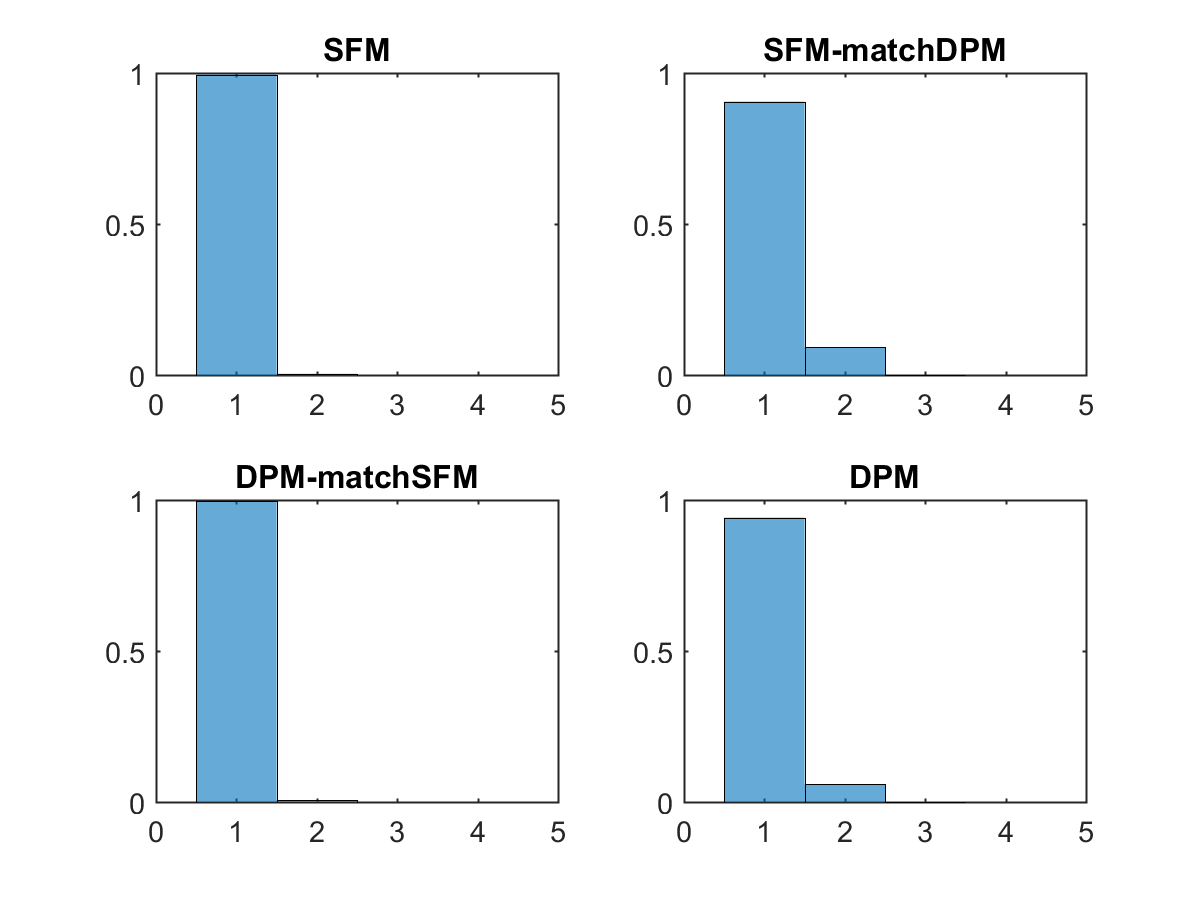}}\\
 \end{tabular}
  \end{center}
 \caption{\fault ;  posterior distributions  $\Prob{\Kn |\ym}$ of the number of clusters $ \Kn$ for mixture of Possion GLMs (left hand side) as well as
 mixtures of  negative binomial GLMs (right hand side); top: based on sparse finite mixtures (SFM),
bottom: based on  Dirichlet process mixtures (DPM) under various hyper priors.}\label{fault_postK} \end{figure}


Table~\ref{tab:fault-bic} and Figure~\ref{fault_postK}  
compare the various   posterior distributions  $\Prob{\Kn |\ym}$ of the number of clusters $ \Kn$ under various hyper priors for both model classes.
  For  mixtures of Poisson GLMs, $K=2$ is selected by the marginal likelihood
  and   $\Knhat=2$,  both  for  sparse finite mixture as well as DPM,   which confirms results
  obtained by \citet{ait:gensc} and \citet{mcl-pee:fin} using alternative methods of model selection.
 For  the more flexible mixture of GLMs based on the negative binomial distribution $K=1$ is selected by the marginal likelihood.
 Also  sparse finite mixtures as well as DPM of GLMs based on the negative binomial distribution estimate
   $\Knhat=1$  cluster. This illustrates that sparse finite mixtures are also  useful for  testing  homogeneity within a Bayesian framework.

One advantage of the  marginal likelihood over sparse finite mixtures and DPMs, however,
     is the possibility  to select the number of clusters and  the appropriate clustering kernel at the same time.
   The model with the largest marginal likelihood in Table~\ref{fault_postK} is the   negative binomial distribution with $K=1$.


\begin{table}[t]
{ \small
\begin{center}
\begin{tabular}{lccccccc}
\hline
$\Prob{\Kn  |\ym}$   & $\Kn=1$ & $\Kn=2$ &$\Kn=3$ &$\Kn=4$ &$\Kn=5$ &$\Kn=6$ &$\Kn \geq 7$  \\
   \hline
Skew normal  & & & & & & & \\
 \hspace*{3mm}   SFM   & & & & & & & \\
  \hspace*{6mm}  $  \ed{0} \sim \Gammad{1 ,200}$  &  0.0127 & \textbf{ 0.760} &  0.193 &  0.029 &  0.005 &  0.000 & 0.000 \\
 \hspace*{6mm}  matched to   DPM & 0.000 &  0.268 & \textbf{ 0.309} &  0.228 &  0.119 &  0.049 & 0.026 \\  
 \hspace*{3mm}  DPM  & & & & & & & \\
 \hspace*{6mm} $\alphaDP \sim \Gammad{2,4}$  & 0.000 &  0.181 & \textbf{ 0.302} &  0.214 &  0.139 &  0.083 & 0.082 \\ 
\hspace*{6mm}   matched to SFM  &0.000 &  \textbf{0.784} &  0.182 &  0.029 &  0.004 &  0.000 &  0.000 \\
 \hline
Skew-$t$  & & & & & & & \\
 \hspace*{3mm}   SFM    & & & & & & & \\
  \hspace*{6mm}  $  \ed{0} \sim \Gammad{1 ,200}$  & 0.263 &  \textbf{0.597} &  0.124 &  0.015 &  0.001 &  0.000 &0.000\\
 \hspace*{6mm}  matched to   DPM & 0.034 &  0.301 &  \textbf{0.320} &  0.205 &  0.094 &  0.032 & 0.013  \\
  \hspace*{3mm} DPM   & & & & & & &   \\
 \hspace*{6mm} $\alphaDP \sim \Gammad{2,4}$  & 0.003 &  \textbf{0.290} &  0.275 &  0.206 &  0.124 &  0.058 &  0.045 \\
\hspace*{6mm}   matched to SFM  & 0.211 &  \textbf{0.492} &  0.214 &  0.065 &  0.016 &  0.002 &  0.000 \\
  \hline
  \end{tabular}\\ [1mm]
  \begin{tabular}{lcccccc}
   \hline
$\log \hat{p}(\ym|K)$  &  & $K=1$ & $K=2$ &$K=3$ &$K=4$ &$K=5$ \\
 \hline
Skew normal   &  FM  ($e_0=4$) &   -689.62 & {\bf  -682.37} &  -684.45 &  -690.41 &  -696.12 \\
Skew-$t$  &   FM  ($e_0=4$)  &  -692.29 & {\bf  -688.98}  &  -690.31 &  -694.11 &  -699.85  \\ 
  \hline
\end{tabular}
\caption{\alz ; the rows in the upper table  show  the posterior distribution  $\Prob{\Kn |\ym}$ of the number of clusters $ \Kn$ for
following mixtures   of univariate skew normal and skew-$t$ distributions: sparse finite mixtures
 with  $K=10$ (SFM) with  hyper priors  $e_0\sim \cG(1,200)$ and $e_0\sim \cG(2,4 K)$  (matched to DPM),
   DPM  with hyper priors $\alphaDP \sim \cG(2,4)$  and $\alphaDP \sim \cG(1,200/K)$ (matched to SFM). The lower  table shows
  log marginal likelihoods, $\log \hat{p}  (\ym|K)$,  estimated for    finite  mixtures with $e_0=4$ (FM)  for  increasing  $K$.}

\label{tab:alz-bic}
\end{center}
}
\end{table}

\subsection{Application to the  \alz\ } \label{secalz}
 Alzheimer disease    is a complex disease that has multiple genetic as well as environmental risk factors. It is commonly
characterized by loss of a wide range of cognitive abilities with aging. For  illustration, data modelled
in  \citet{fru-pyn:bay} through (standard) finite mixtures of  skew normal and skew-$t$ distributions are reanalyzed.
The data set consists of   $N=451$  subjects,  whose level of cognition was clinically evaluated proximate to their death
 based on tests of cognitive functions and summarized by a mean global cognition score, with higher scores suggesting better
 cognitive capabilities; see \citet{ben-etal:rus} for more details on the corresponding study. The true number of  groups in these data is equal to two.
The goal of the exercise is to investigate, if sparse finite mixtures with non-Gaussian components based on parametric densities  such as
univariate skew normal and skew-$t$ distributions are able to detect the true number of clusters and to compare them to DPM models.

\begin{figure}[t]\begin{center}
\begin{tabular}{cc}
\scalebox{0.3}{\includegraphics{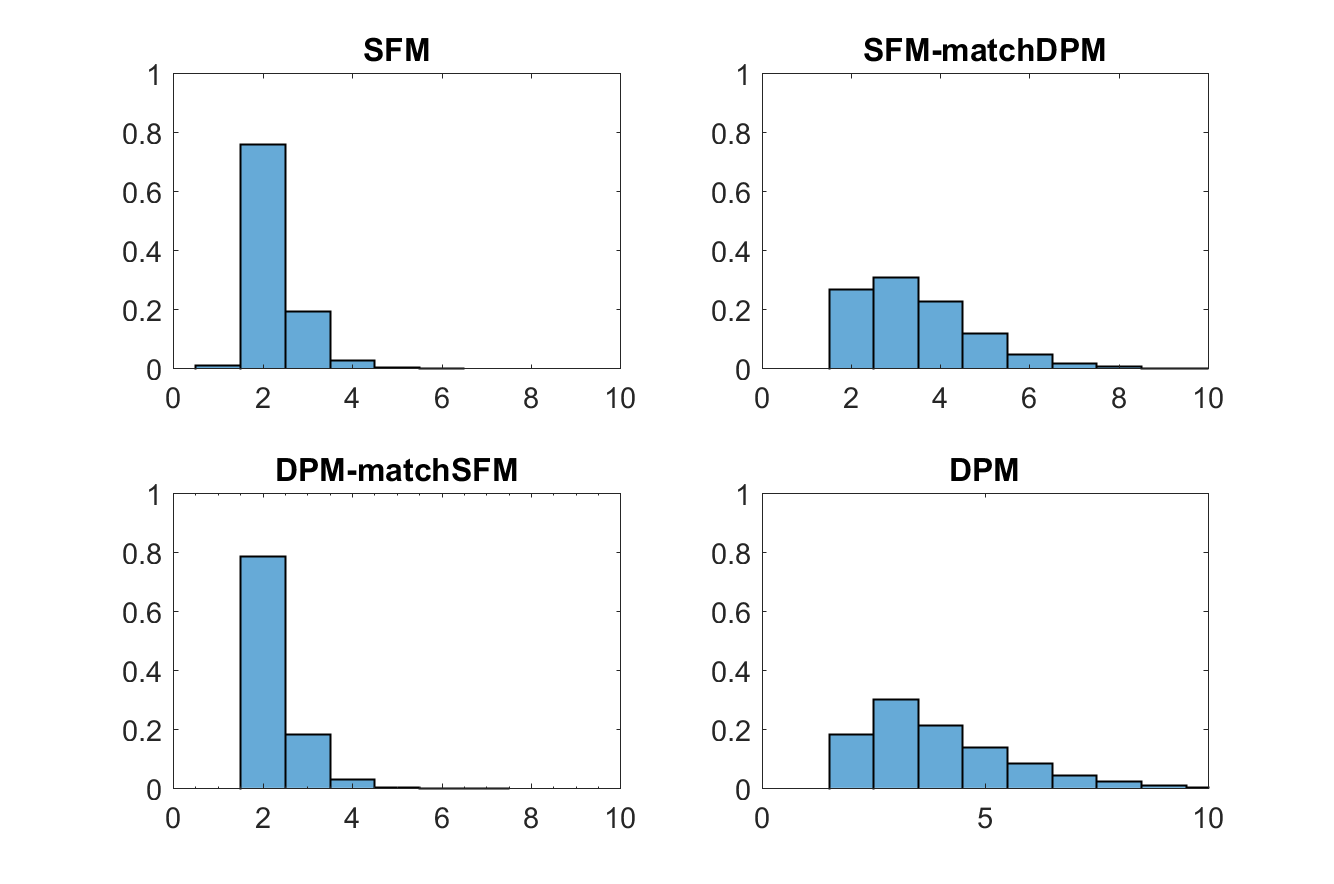}} \hspace*{0.5cm}  &  \hspace*{0.5cm}
 \scalebox{0.3}{\includegraphics{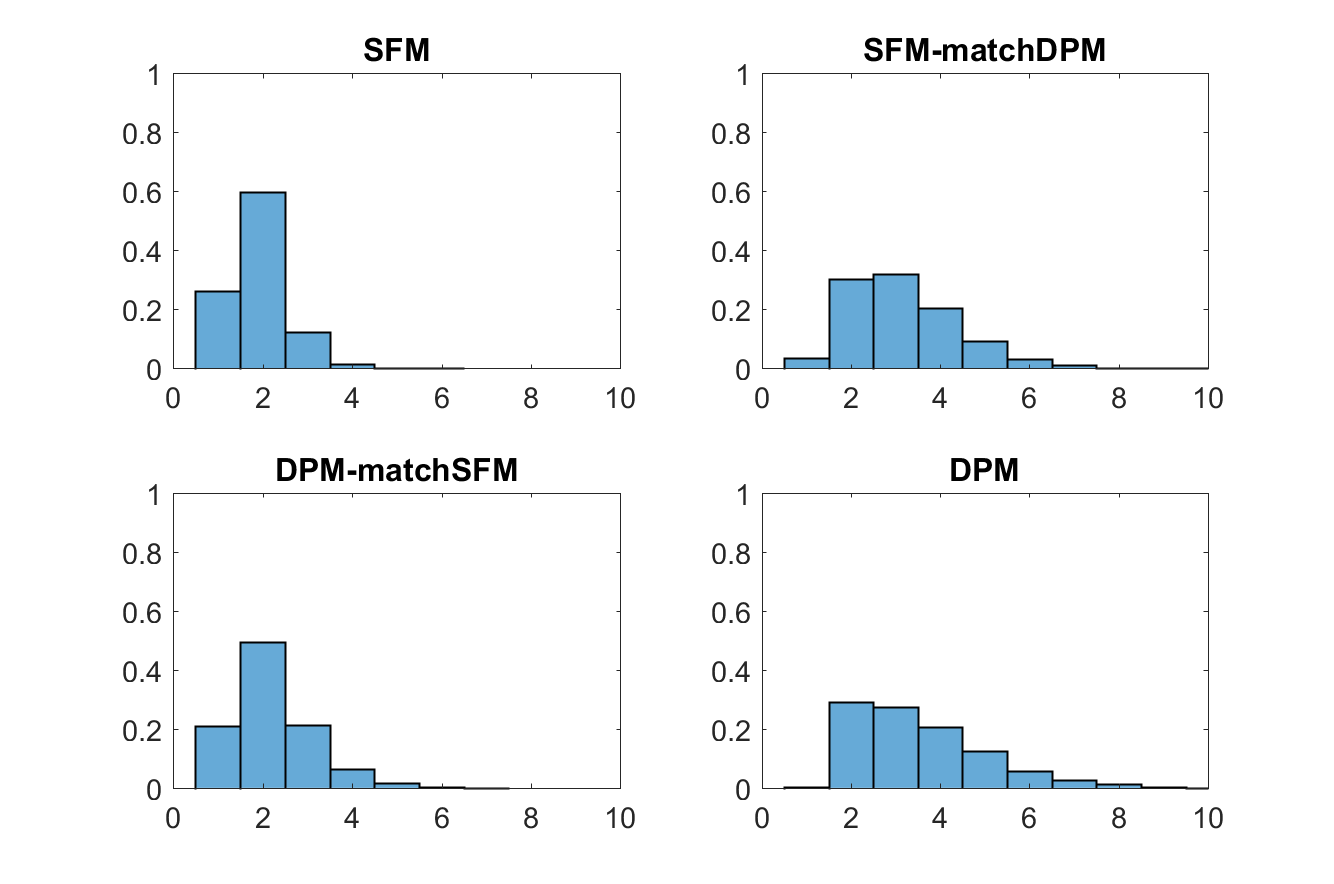}}\\
 \end{tabular}
  \end{center}
 \caption{\alz ;  posterior distributions  $\Prob{\Kn |\ym}$ of the number of clusters $ \Kn$ for mixtures of skew normal (left hand panel) as well as
 mixtures of  skew-$t$  distributions (right hand panel);   top row in each panel:  sparse finite mixtures  with  $K=10$,   $e_0\sim \cG(1,200)$ (left column) and  matched prior $e_0\sim \cG(2,4 K)$ (right column);
 bottom  row in each panel: DPM with $\alphaDP \sim \cG(2,4)$ (right column) and  matched prior  $\alphaDP \sim \cG(1,200/K)$ (left column).}\label{fig:alz} \end{figure}

\citet{fru-pyn:bay} considered various methods for selecting $K$  for   skew normal  and   skew-$t$ mixtures under the prior $e_0=4$.
 In particular,  DIC criteria  \citep{cel-etal:dev}  turned out
 to be extremely sensitive to prior  choices for  the cluster-specific parameter $(\xis_k ,\alpha_k, \omegas_k )$.
The marginal likelihoods  of a standard finite mixture model with   $\ed{0}=4$ are
compared in  Table~\ref{tab:alz-bic} 
  to sparse finite  skew normal and skew-$t$ mixture models, where
$K=10$ and  $\ed{0} \sim \Gammad{1 ,200}$, as well as to DPMs of these same type.
 Table~\ref{tab:alz-bic} and Figure~\ref{fig:alz}  summarize the posterior distributions   $\Prob{\Kn |\ym}$ of the number of clusters $ \Kn$ under various hyper priors.

%

 Again,  Figure~\ref{fig:alz} illustrates that the main difference between the resulting posterior distributions of  $ \Kn$ is not wether a Dirichlet process mixtures
or a finite mixture model is applied. Rather, the apparent difference is due to changes in the hyper prior. A  sparse  prior on the precision  parameters
$\ed{0}$ and $\alphaDP$ yields a clear decision concerning  $ \Kn$, namely  selecting  $\Knhat=2$ for both types of  clustering kernels.
This is true
both for a sparse finite mixture and  a \lq\lq sparse\rq\rq\ DPM where  the hyper prior for $\alphaDP$ is  matched  to the sparse finite mixture.
However,  for a prior that does not force sparsity, both  sparse finite mixtures as well as DPM  overestimate the number of clusters
   with $\Knhat=3$ for the skew normal distribution and are more or less undecided between two and three clusters for the  skew-$t$ mixture.

\begin{table}[t]
	{ \small
		\begin{center}
			\begin{tabular}{lccc}
				\hline
				  & $\Knhat$ & $\Ew{\ed{0}|\ym}$  & $\Ew{\alphaDP |\ym}$ \\
				\hline
				Skew normal  & & & \\
				\hspace*{3mm}   SFM  & & & \\
				\hspace*{6mm}  $  \ed{0} \sim \Gammad{1 ,100}$  &  15 &  0.089(0.04, 0.14)  &   \\
				\hspace*{6mm}  matched to   DPM & 14 &    0.094(0.04,0.15)  &  \\
				\hspace*{3mm}  DPM  & & &   \\
				\hspace*{6mm} $\alphaDP \sim \Gammad{2,4}$  & 26 & &  1.71(0.99,2.49)  \\
				\hspace*{6mm}   matched to SFM  &23 &  & 0.68(0.38,0.98)  \\
				\hline
				Skew-$t$  & & &\\
				\hspace*{3mm}   SFM   & &&  \\
				\hspace*{6mm}  $  \ed{0} \sim \Gammad{1 ,100}$  & 11 &   0.058(0.03,0.10)  & \\
				\hspace*{6mm}  matched to   DPM & 10 &   0.067(0.03, 0.11) & \\
				\hspace*{3mm} DPM   & &  & \\
				\hspace*{6mm} $\alphaDP \sim \Gammad{2,4}$  & 14 &  &  1.20(0.56,1.86)   \\
				\hspace*{6mm}   matched to SFM  & 10 &  &  0.37(0.15,0.59) \\
				\hline
			\end{tabular}\\ [1mm]
		\begin{tabular}{lcccccc}
			\hline
			$\log \hat{p}(\ym|K)$  &  
			& $K=2$ &$K=3$ &$K=4$ &$K=5$ &$K=6$  \\
			\hline
			%
			Skew normal  & FM ($e_0=4$)   & 
			-19160 &  -19116 &  -18818 &  -18388 &  -18045  \\  
			Skew-$t$  & FM ($e_0=4$)
			&             
			-18980 &  -18433 &  -18131 &  -17918 &  -17915   \\   
			\hline
		\end{tabular}
		\caption{\DLBCL ;  estimated  number of clusters $ \Knhat$  for  following mixtures   of multivariate skew normal and skew-$t$ distributions:
			sparse finite mixtures  with  $K=20$ (SFM) with  hyper priors  $e_0\sim \cG(1,100)$ and $e_0\sim \cG(2,4 K)$  (matched to DPM),
			DPM  with hyper priors $\alphaDP \sim \cG(2,4)$  and $\alphaDP \sim \cG(1,100/K)$ (matched to SFM).
The lower  table shows
			log marginal likelihoods, $\log \hat{p}  (\ym|K)$,  estimated for    finite  mixtures with $e_0=4$ (FM)  for  increasing  $K$.}\label{tab:DLBCL}
		\end{center}
	}
\end{table}

\begin{figure}[t]\begin{center}
\begin{tabular}{cc}
 \scalebox{0.3}{\includegraphics{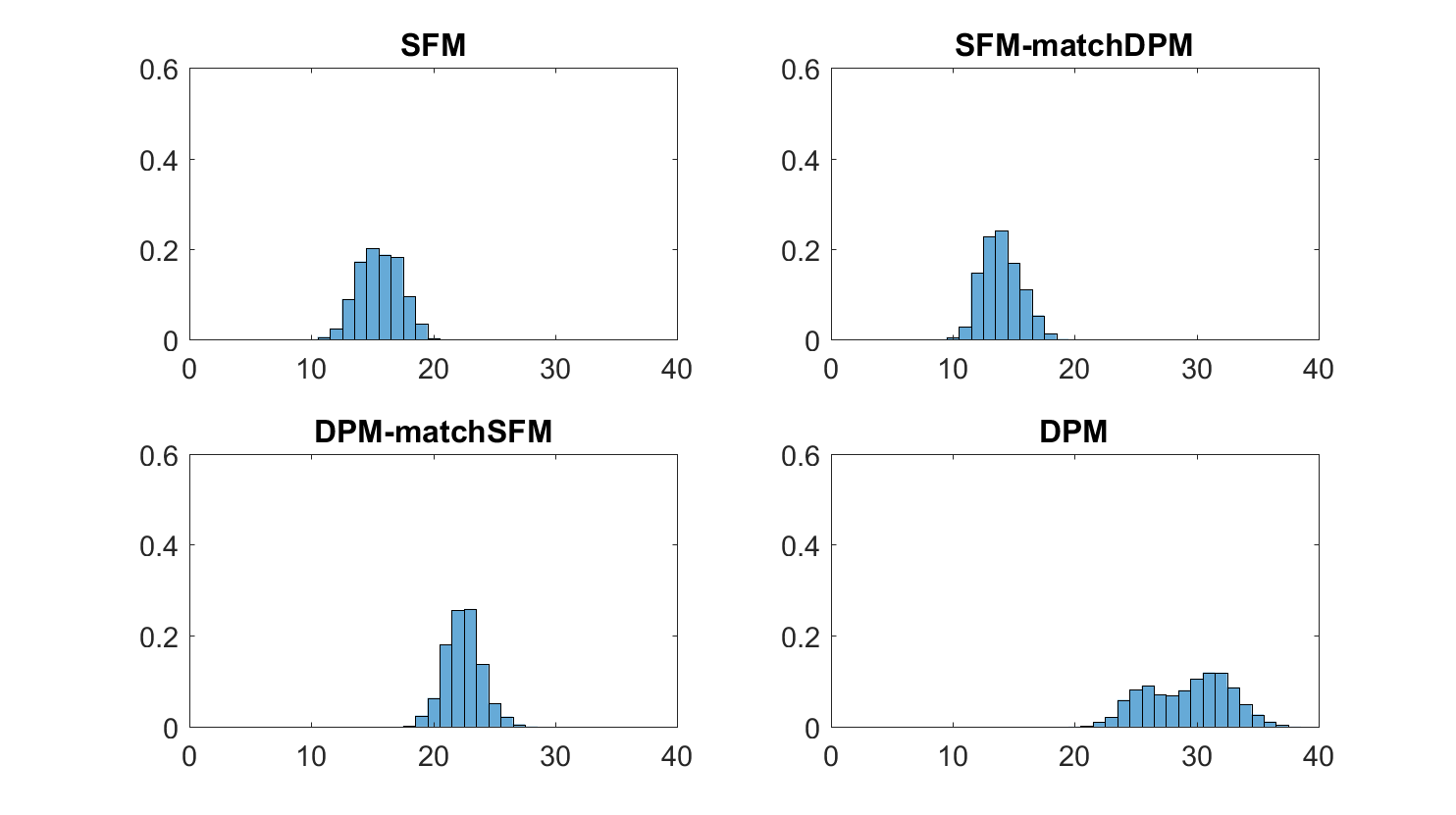}}  \hspace*{0.5cm}  &  \hspace*{0.5cm}
  \scalebox{0.3}{\includegraphics{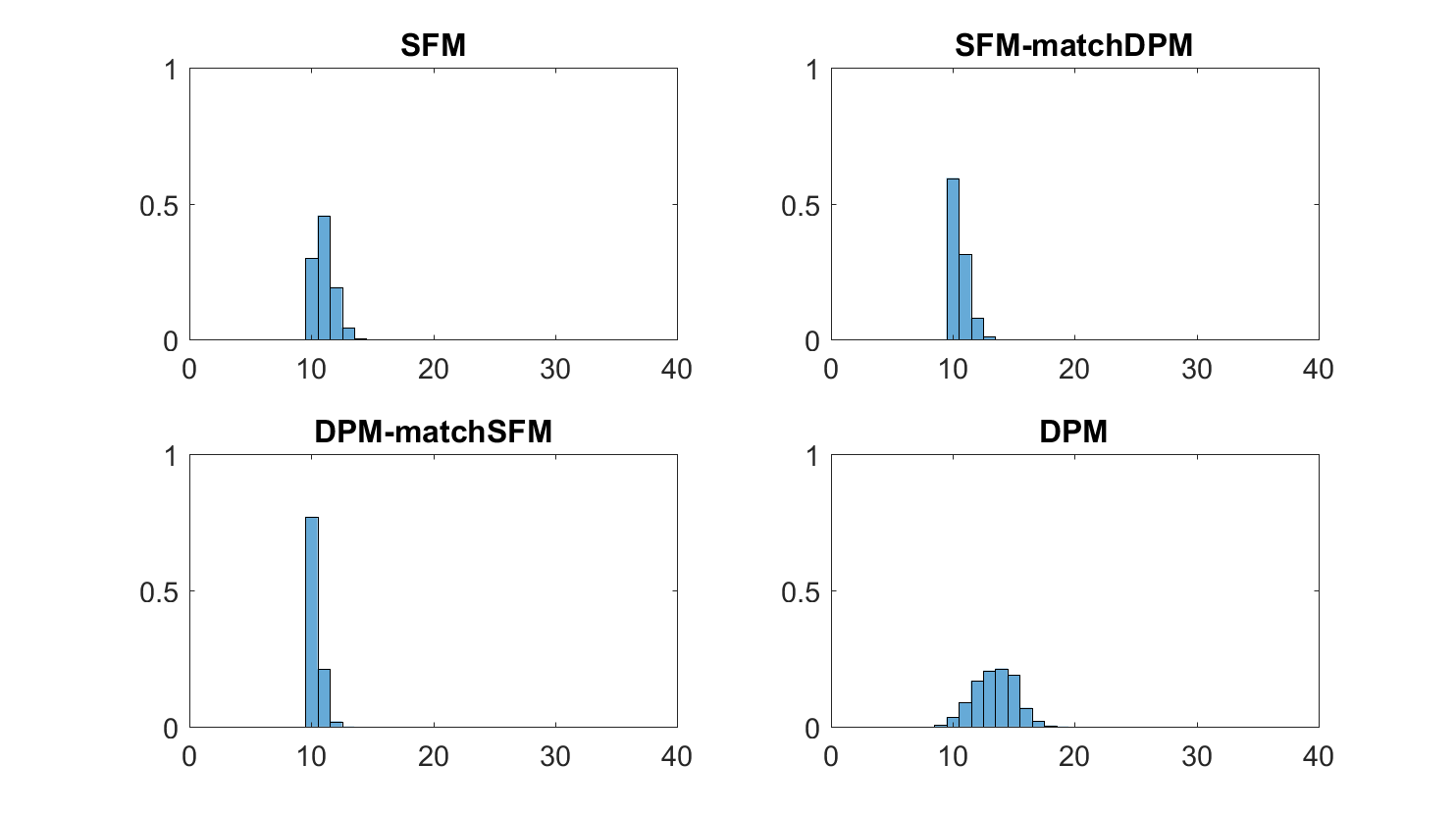}}\\
 \end{tabular}
\end{center}
 \caption{\DLBCL ;  posterior distributions  $\Prob{\Kn |\ym}$ of the number of clusters $ \Kn$ for
 mixtures of skew normal (left hand panel) as well as
 mixtures of  skew-$t$  distributions (right hand panel);   top row in each panel:  sparse finite mixtures  with  $K=20$,   $e_0\sim \cG(1,100)$ (left column) and  matched prior $e_0\sim \cG(2,4 K)$ (right column);
 bottom  row in each panel: DPM with $\alphaDP \sim \cG(2,4)$ (right column) and  matched prior  $\alphaDP \sim \cG(1,100/K)$ (left column).}\label{fig:DLBCL} \end{figure}

  The choices obtained from both sparse finite mixture models and DPM coincide  with  the decision obtained by the marginal likelihood.
An advantage of the  marginal likelihood over sparse mixtures is that, in addition to $K$, the clustering kernel can be selected.
For the data at hand,  finite mixtures of skew normal distributions are preferred to  skew-$t$ distributions.

%

\subsection{Applications to flow cytometric data}  \label{secDLBLC}
%

 To assess how sparse finite mixtures scale to larger data sets,
an  application to   flow cytometry data  is investigated. The three-dimensional
DLBCL data set \citep{lee-mcl:mod} consists of  $N=7932$
observations,   with class labels which were determined manually. The true number of  groups in these data is equal to 4.
 \citet{mal-etal:ide} fitted  a sparse finite mixture-of-mixtures model to these data with  $K=30$ and $e_0=0.001$. The component densities
were estimated in a semi-parametric manner through a Gaussian mixture with $ L=15$ components and inference  identifies  $ \Knhat =4$
such non-Gaussian clusters. The resulting error rate (0.03) outperformed the error rate of 0.056 reported by \cite{lee-mcl:mod}. 

The goal of this application  is to investigate,  whether sparse finite mixtures with non-Gaussian components  based on parametric densities  such as
the multivariate skew normal and skew-$t$  distributions are able to detect this true number of clusters.
Sparse finite mixtures with $K=20$ and $\ed{0} \sim \Gammad{1,100}$,
 as well as DPM of the corresponding type are fitted to these data and results are reported in Table~\ref{tab:DLBCL} and Figure~\ref{fig:DLBCL}. As it turns out, the posterior
 expectation   of both precision parameters, i.e. $\Ew{\alphaDP|\ym}$ as well as  $\Ew{\ed{0}|\ym}$ are  pretty large,
indicating that a lot of components are needed to describe these data.
Consequently,  the estimated number of clusters  $\hat{K}_+$  is much larger than four   for any of these mixtures.
This finding is confirmed by the marginal likelihoods.
Obviously, neither  skew normal nor skew-$t$ distributions are as flexible as the  mixture-of-mixtures model
introduced by \citet{mal-etal:ide} to capture departure from normality  for these data.

%
%

\begin{table}
\begin{center}
\begin{tabular}{lrrrcccc}
\hline & &  & & \multicolumn{2}{c}{SFM} &   \multicolumn{2}{c}{RM}\\
Data  set & $N$ & $r$ & $d$ &  $\Ew{\ed{0}|\ym}$ & 95\%~CI  & $\Ew{\ed{0}|\ym}$ & 95\%~CI  \\  
   \hline
 \eye\  &  101 & 1 & 1 &  0.020 & (0.004, 0.04) & 0.37 &(0.18,0.5) \\
 \fear\ & 93 & 3  & 7 & 0.010 & (0.0007,0.023) & 1.30 &(0.09,3.01)\\
 \fault\ (NegBin) & 32 & 1 & 3 &  0.004 &(0,0.014) & 0.04 &(0,0.13)\\
\alz\ (SkewN) & 451 & 1 & 3 & 0.009& (0.0001,0.022) & 0.36& (0.18,0.5) \\
  \hline
\end{tabular}
\caption{Posterior expectations $\Ew{\ed{0}|\ym}$ of  $\ed{0}$ together with  95\% confidence regions  for the various data sets;
sparse finite mixture  with $K=10$ and  $ \ed{0} \sim  \Gammad{1 ,200} $  (SFM) versus overfitting mixtures with $K=10$ and
$ \ed{0} \sim \Uniform{0, d/2}$ (RM).}\label{tab_e0}
\end{center}
\end{table}

\begin{figure}[h]
\begin{center}
\scalebox{0.4}{\includegraphics{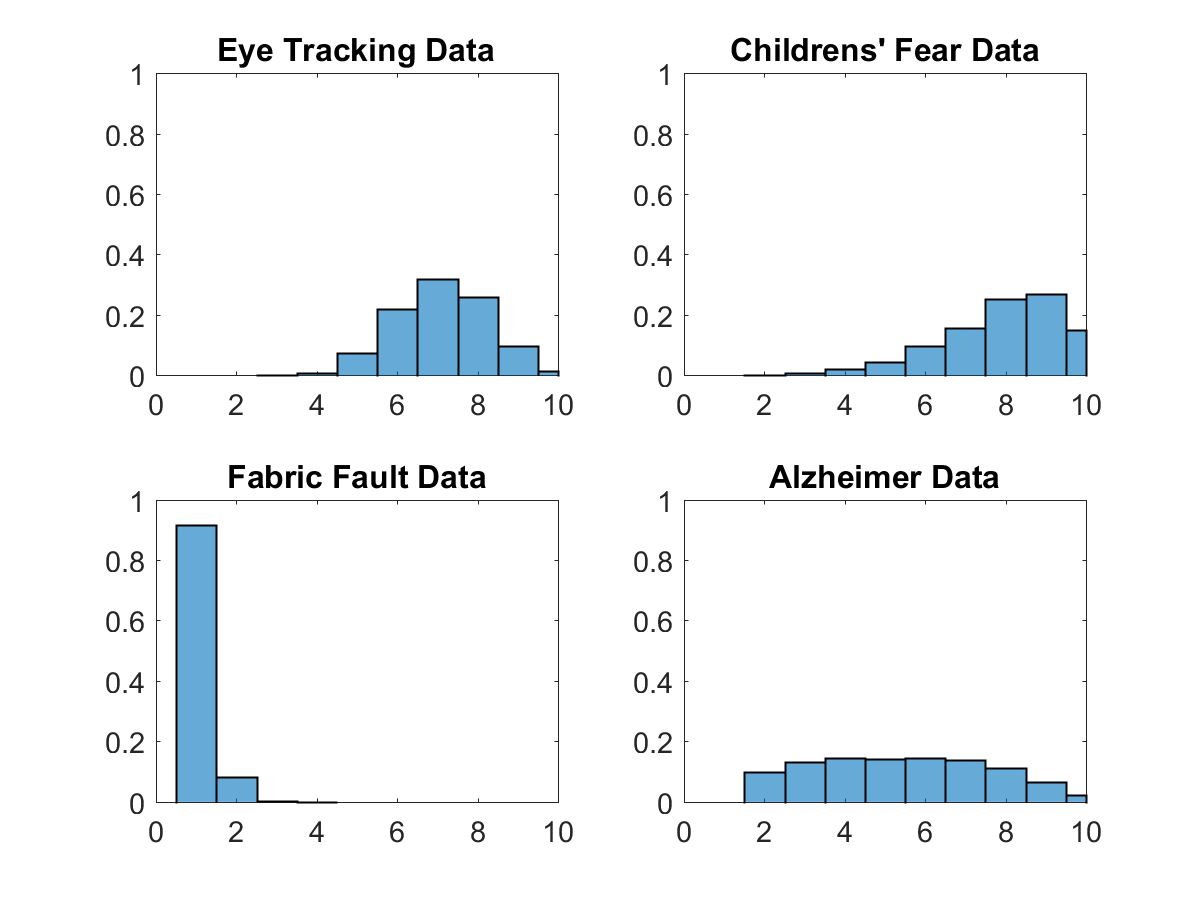}}
  \end{center}
 \caption{Posterior distributions  $\Prob{\Kn |\ym}$ of the number of clusters $ \Kn$ for the various data sets for a sparse finite mixture
 with $K=10$ and  prior $ \ed{0} \sim \Uniform{0, d/2}$  derived from the criterion of \citet{rou-men:asy}.}\label{fig_menrou}
 \end{figure}

\section{Discussion and concluding remarks}\label{sec:dis}

This paper  extends the concept of sparse finite mixture models, introduced by \citet{mal-etal:mod} for Gaussian clustering kernels,
to a wide range of   non-Gaussian mixture models, including  Poisson mixtures,
latent class analysis, mixtures of GLMs, skew normal and skew-$t$ distributions.
 Opposed to  common belief,  this paper shows that finite mixture models do not necessarily assume that
the  number   of clusters   is known.
 As exemplified for several case studies  in Section~\ref{sec:appl},   the  number of clusters was estimated a posteriori from the data and  ranged
 from  $ \Knhat=1$ (for the \fault\  under  a mixture of negative binomial  GLMs)
to   $ \Knhat=4$  (for the \eye ), when sparse finite mixtures with  \g{$K=10$ components} were fitted.

Sparse finite mixture models are  based on overfitting mixture distributions, where the number of  clusters $\Kn$
among $N$ data points  generated from such a mixture is, with high probability, smaller than $K$ a priori.
This is achieved by choosing a symmetric Dirichlet prior on the weight distribution $(\eta_1,\ldots,\eta_K) \sim \Dirinv{K}{\ed{0}}$,
with a sparsity prior on $\ed{0}$  that favours very small values.

A theoretical justification for sparse finite mixture models seems to emerge from an asymptotic results of  \citet{rou-men:asy},
who show that the asymptotic behaviour of the mixture posterior  $p( \thetav_1, \ldots, \thetav_K,\etav| \ym_1, \ldots,\ym_N)$
as $N$ goes to infinity is determined by the hyperparameter  $\ed{0}$ of the symmetric Dirichlet prior 
$\Dirinv{K}{\ed{0}}$.
 Let $d=\dim{\thetav_k}$ be the dimension of the  component-specific  parameter $\thetav_k$ in a  mixture distribution (\ref{mix:dist}) with
 $\Ktrue$   distinct  components (i.e. $\thetav_k \neq \thetav_l$, $k\neq l$) with non-zero weights.
 If  $ \ed{0} < d/2$,  then  the posterior  distribution   of an overfitting   mixture distribution with $K>\Ktrue$ components  asymptotically  concentrates
 over regions forcing the sum of the weights of  the $K-\Ktrue$  extra components to concentrate at 0.
%
%
 Hence,  if  $ \ed{0} < d/2$,  all  superfluous components in an overfitting mixture are  emptied,
   as the number of observations $N$ goes to infinity.
 However,  the implications of  this  important result  
 for the posterior concentration  of  the number of data clusters   $\Kn$
 are  still unclear.   As shown by \citet{mil-har:sim},
 the number of  clusters  $\Kn$ in data generated from a finite mixture distribution of order $\Ktrue$  converges  to $\Ktrue$, as $N$  goes to infinity, if
 $K=\Ktrue$.
Conditions under which  such a convergence holds,  if   $\Ktrue$  is unknown and  an overfitting mixture with $K>\Ktrue$ is fitted,  are an interesting
venue of  future research.


As noted by \citet{mal-etal:mod},  who applied overfitting Gaussian mixtures to model-based clustering of quite a few benchmark data sets,
  values of  $\ed{0}$ much smaller  than  \citet{rou-men:asy}'s  threshold $ d/2$   are needed in practice to identify the right number of clusters.
  We obtained similar results for the extensions and applications considered in the present paper.
Table~\ref{tab_e0} summarizes the posterior expectations $\Ew{\ed{0}|\ym}$ as well as 95\% confidence regions of $e_0$
for   various  data sets fitted in Section~\ref{sec:appl} under the
sparse prior $e_0\sim\cG(1,200)$, with prior expectation $\Ew{e_0}=0.005$.    These results confirm that  the posterior distribution of  $e_0$
is concentrated over values that are
 considerably smaller than $d/2$ (the dimensions $d$ are also reported in the table).
  To see,  whether the data alone would have been  informative about  $e_0$ for these case studies, the  uniform prior   $ \ed{0} \sim \Uniform{0, d/2}$ over
 the region $[0, d/2]$ is considered. 
%
 The corresponding  posterior expectations $\Ew{\ed{0}|\ym}$,   reported in Table~\ref{tab_e0},
are considerably larger than for the sparsity prior.  As  can be seen in Figure~\ref{fig_menrou},
 this leads  to posterior distributions $p(\Kn|\ym)$  that overfit  the number of clusters for all data sets
 considerably, except for the homogeneous \fault . These results indicate that   regularisation of the posterior distribution through a sparsity prior
that encourages values of  $\ed{0}$ much smaller than $d/2$ is essential for identifying the number of clusters.

Introducing a sparsity prior avoids overfitting the number of clusters not only for  finite mixtures, but also (somewhat unexpectedly)
for Dirichlet process mixtures which are known to overfit the number of  clusters \citep{mil-har:sim}.
\comment{For the data  considered in the present paper, overfitting could
 be avoided  through a prior on the precision parameter $\alphaDP$ that encouraged very small values.
 When matching  the priors of $\ed{0}$  in sparse finite mixtures  and  $\alphaDP$  in DPM,  the posterior distribution of the
number of clusters was more influenced by these hyper priors  than whether the
 mixture was finite or infinite.  It would be interesting to investigate, if  this proximity of both model classes also holds more generally.}

 \comment{Another avenues for future research concern MCMC estimation. Although we did not encounter problems with full conditional Gibbs sampling for
 our case studies,  more efficient algorithms could be designed  by using  parallel tempering
 as in \citet{van-etal:ove}  or by exploiting ideas from BNP (e.g. \citet{fal-bar:gib}).}



\subsection*{Acknowledgements}

This research  was  funded by the Austrian Science Fund (FWF): P28740. We owe special thanks to Bettina Gr\"un for  many helpful comments on preliminary versions of this paper.




\bibliographystyle{chicago}
\bibliography{sylvia_kyoto,MixofMix}

\end{document}